\documentclass[traditabstract]{aa}
\pdfoutput=1
\usepackage{verbatim} 
\usepackage{amsmath}
\usepackage{amssymb}
\usepackage{graphicx}
\usepackage{booktabs}
\usepackage{natbib}
\usepackage[colorlinks=true,linkcolor=blue,urlcolor=blue,citecolor=blue]{hyperref}
\usepackage{xspace}
\usepackage{color}

\usepackage{txfonts}
\usepackage{twoopt}
\bibpunct{(}{)}{;}{a}{}{,} 
\newcommand{\msun}{{\rm M}_{\odot}}
\newcommand{\lsun}{{\rm L}_{\odot}}

\newcommand{\bonnsai}{\mbox{\textsc{Bonnsai}}\xspace}
\newcommand{\fastwind}{\mbox{\textsc{Fastwind}}\xspace}
\newcommand{\gaia}{\mbox{Gaia}\xspace}
\newcommand*{\eg}{e.g.\@\xspace}
\newcommand*{\ie}{i.e.\@\xspace}
\newcommand{\cov}{\vec{\Sigma}}
\newcommand{\rot}{\vec{R}}
\titlerunning{Correlations in Bayesian methods}
\authorrunning{Fabian~R.N.~Schneider~et~al.}

\begin{document}

\title{BONNSAI: correlated stellar observables in Bayesian methods}

\author{F.R.N.~Schneider\inst{\ref{OXFORD}}\thanks{fabian.schneider@physics.ox.ac.uk}
\and N.~Castro\inst{\ref{AIFA}}
\and L.~Fossati\inst{\ref{GRAZ}}
\and N.~Langer\inst{\ref{AIFA}}
\and A.~de Koter\inst{\ref{AMSTERDAM},\ref{LEUVEN}}}

\institute{Department of Physics, University of Oxford, Denys Wilkinson Building, Keble Road, Oxford OX1 3RH, United Kingdom\label{OXFORD}
\and Argelander-Institut f{\"u}r Astronomie der Universit{\"a}t Bonn, Auf dem H{\"u}gel~71, 53121~Bonn, Germany\label{AIFA} 
\and Space Research Institute, Austrian Academy of Sciences, Schmiedlstrasse 6, 8042 Graz, Austria\label{GRAZ}
\and Astronomical Institute 'Anton Pannekoek', Amsterdam University, Science Park 904, 1098 XH, Amsterdam, The Netherlands\label{AMSTERDAM} 
\and Instituut voor Sterrenkunde, KU Leuven, Celestijnenlaan 200D, 3001, Leuven, Belgium\label{LEUVEN}}

\date{Received day month 2016 / Accepted day month 2016}

\abstract{In an era of large spectroscopic surveys of stars and big data, sophisticated statistical methods become more and more important in order to infer fundamental stellar parameters such as mass and age. Bayesian techniques are powerful methods because they can match all available observables simultaneously to stellar models while taking prior knowledge properly into account. However, in most cases it is assumed that observables are uncorrelated which is generally not the case. Here, we include correlations in the Bayesian code \bonnsai by incorporating the covariance matrix in the likelihood function. We derive a parametrisation of the covariance matrix that, in addition to classical uncertainties, only requires the specification of a correlation parameter that describes how observables co-vary. Our correlation parameter depends purely on the method with which observables have been determined and can be analytically derived in some cases. This approach therefore has the advantage that correlations can be accounted for even if information for them are not available in specific cases but are known in general. Because the new likelihood model is a better approximation of the data, the reliability and robustness of the inferred parameters are improved. We find that neglecting correlations biases the most likely values of inferred stellar parameters and affects the precision with which these parameters can be determined. The importance of these biases depends on the strength of the correlations and the uncertainties. For example, we apply our technique to massive OB stars, but emphasise that it is valid for any type of stars. For effective temperatures and surface gravities determined from atmosphere modelling, we find that masses can be underestimated on average by $0.5\sigma$ and mass uncertainties overestimated by a factor of about 2 when neglecting correlations. At the same time, the age precisions are underestimated over a wide range of stellar parameters. We conclude that accounting for correlations is essential in order to derive reliable stellar parameters including robust uncertainties and will be vital when entering an era of precision stellar astrophysics thanks to the \gaia satellite.}


\maketitle

\section{Introduction}\label{sec:introduction}

With the advent of large stellar spectroscopic surveys, powerful telescopes, and advanced spectral modelling capabilities, more and more is known about individual stars with ever increasing accuracy and precision. In particular, the \gaia satellite will revolutionise the precision with which distances, luminosities, and other stellar parameters can be determined. 

Accuracy and precision are essential to many astrophysical applications, for example when determining fundamental properties of exoplanets and the architectures of planetary systems from inferred masses, radii, and ages of their host stars \citep[\eg][]{2016A&A...585A...5B}; when studying the dynamical and chemical evolution of the Galaxy from F and G stars \citep[\eg][]{2014MNRAS.438.2753M}; when testing stellar models with in-depth observations of binary stars \citep[\eg][]{2002A&A...396..551L,2010A&ARv..18...67T}; or when investigating the consistency of different stellar age estimate methods \citep[\eg][]{2015A&A...577A..90M}. Also, in statistical studies of large samples of stars, systematic biases have the potential to hamper our interpretation of the data \citep[\eg][]{2005A&A...436..127J,Fossati+2016}.

In order to determine robust and reliable stellar parameters, sophisticated statistical methods are indispensable and systematic uncertainties need to be understood. This includes systematics in the stellar models \citep[\eg][]{2013A&A...560A..16M,2015MNRAS.447.3115J,2015A&A...575A.117S,2016A&A...586A.119S}, but also systematics in the statistical techniques used to compare observations of stars with stellar models. Bayesian statistical methods \citep[\eg][]{2005A&A...436..127J,2006A&A...458..609D,2007ApJS..168..297T,2007MNRAS.377..120S,2009AnApS...3..117V,2010MNRAS.407..339B,2013MNRAS.429.3645S,2014MNRAS.443..698S,2014A&A...570A..66S,2015A&A...575A..36M} have proven to be powerful because they can compare all available observables including error bars simultaneously to stellar models; they also take prior knowledge such as initial mass functions properly into account. Moreover, such methods have the potential to deliver robust and reliable error bars of inferred stellar parameters that do not suffer from biases, for example due to neglecting the time spent by stars in various regions of the Hertzsprung--Russell (HR) diagram \citep[\eg][]{2004MNRAS.351..487P}. 

For simplicity, it is in most cases assumed that observables are independent of each other, \ie that they are uncorrelated. However, this assumption is not generally true and many observables are in fact correlated. For example, effective temperatures and surface gravities are correlated if inferred from stellar atmosphere modelling because they can influence diagnostic lines in a similar way, \ie effective temperature and surface gravity depend on each other. The same is true when converting observed effective temperatures, distances, and magnitudes of stars into bolometric luminosities. To accomplish this it is necessary to use the bolometric correction, which is a function of effective temperature, because of the approximate black-body behaviour of stars. Therefore, the luminosity also depends on effective temperature: bolometric luminosity and effective temperature co-vary, \ie they are correlated.

Correlations are valuable information and, when not accounted for, can bias the comparison of observations with stellar models. Correlations change the likelihood function and, as we show in this paper, thereby the precision and most likely values of inferred fundamental stellar parameters such as mass and age. They provide additional information on the observables that ultimately result in more reliable stellar parameters. Taking correlations properly into account may decrease the error bars such that stellar parameters can be inferred with higher precision.

\bonnsai\footnote{The BONNSAI web-service is available at \href{http://www.astro.uni-bonn.de/stars/bonnsai}{www.astro.uni-bonn.de/stars/bonnsai}.} \citep{2014A&A...570A..66S} is a Bayesian statistical method that matches all the available observables of stars simultaneously to models of stellar evolution in order to determine fundamental stellar parameters such as mass and age, to predict yet unobserved stellar parameters, and to probe theses models using sophisticated goodness-of-fit tests. Being a Bayesian method, it takes prior knowledge such as initial mass functions properly into account. In this paper, we generalise the likelihood function of \bonnsai by adding the covariance matrix to be able to account for correlated stellar observables. So far, correlations of stellar parameters are typically not reported in the literature, but only marginalised $1\sigma$ uncertainties. In order to facilitate the use of such stellar parameters, we derive a parametrisation of the covariance matrix that depends on conventional $1\sigma$ uncertainties, and also on a correlation parameter that describes how two observables co-vary and that purely depends on the method with which the observables have been determined. Once the correlation parameter for a particular method is known, this parametrisation has the advantage that correlations can be accounted for when matching observables against stellar models even if only marginalised $1\sigma$ error bars are available. The new likelihood model can be applied in various situations (also outside stellar astrophysics) and is valid for any kind of star in any evolutionary state.

In Sec.~\ref{sec:method} we describe the new likelihood model and compare it to the old one and to even more accurate numerical models. We investigate how the new likelihood model affects the precision and most likely values of inferred initial masses and ages in Sec.~\ref{sec:results}, and summarise our conclusions in Sec.~\ref{sec:conclusion}. A version of \bonnsai that includes the new likelihood model is available through a web-interface\footnote{\href{http://www.astro.uni-bonn.de/stars/bonnsai}{www.astro.uni-bonn.de/stars/bonnsai}}.

\section{Method}\label{sec:method}

\subsection{The new likelihood model}\label{sec:new-likelihood-model}

The heart of any Bayesian method such as \bonnsai is Bayes' theorem,
\begin{equation}
p(\vec{m}|\vec{d}) \propto p(\vec{d}|\vec{m}) p(\vec{m}).
\label{eq:bayes-theorem}
\end{equation} 
Bayes' theorem states that the posterior probability distribution $p(\vec{m}|\vec{d})$, \ie the probability distribution of the model parameters $\vec{m}$ given observational data $\vec{d}$, follows from the likelihood $p(\vec{d}|\vec{m})$, \ie the probability distribution of the observational data given the model parameters and the prior distribution $p(\vec{m})$ that encompasses a priori knowledge about the model parameters.

In \bonnsai and similar methods it is usually assumed that the likelihood function can be written as a product of Gaussian functions for each observable. This assumption is true if the observables are normally distributed and uncorrelated. 

If the observables are correlated, the product of independent Gaussian likelihoods is only a zeroth order approximation of the data; this approximation may still be adequate to describe the data if the correlations are weak. In a first-order approximation, it is possible to generalise the Gaussian likelihood function by introducing the covariance matrix $\cov$,
\begin{equation}
p(\vec{d}|\vec{m})=\frac{1}{\sqrt{(2\pi)^k |\cov|}} \exp \left[ -\frac{1}{2} \left( \vec{d} - \vec{d}(\vec{m})\right)^\mathrm{T} \cov^{-1} \left( \vec{d} - \vec{d}(\vec{m})\right) \right],
\label{eq:likelihood}
\end{equation}
where $k$ is the dimension, \ie the number of observables, $|\cov|$ the determinant of the covariance matrix, and $\vec{d}(\vec{m})$ the predicted observables for model parameters $\vec{m}$. If the observables are uncorrelated, the covariance matrix has only diagonal elements and the likelihood function in Eq.~\ref{eq:likelihood} reduces to a product of independent Gaussian functions. The likelihood model in Eq.~\ref{eq:likelihood} may still not be good enough to properly approximate the data, for example if the shape of the likelihood is that of a banana. In that case it is probably easiest to use a numerical model of the likelihood function in Bayes' theorem.

Covariances of inferred stellar parameters are typically not reported in the literature, but only $1\sigma$ uncertainties of individual observables; we call these $1\sigma$ uncertainties ``conventional uncertainties/error bars''. Our aim is therefore to derive a parametrisation of the covariance matrix that is fully specified by conventional error bars and additional parameters that we call ``correlation parameters'' that only depend on the method with which observables have been derived. Once the correlation parameters are known for a given method, this approach will enable us to incorporate correlations even in such cases where only conventional error bars are available.

Covariances are often expressed in terms of Pearson's correlation parameter $\rho_{pq}=\mathrm{Cov}(p,q)/\sigma_p \sigma_q$ where $\mathrm{Cov}$ is the covariance of observables $p$ and $q$, and $\sigma_p$ and $\sigma_q$ are the respective conventional $1\sigma$ uncertainties. With this notation, the covariance matrix $\cov$ has the square of the conventional uncertainties on its diagonal and $\rho_{pq}\sigma_p \sigma_q$ on its off-diagonal (recall that $\cov$ is symmetric). In this way, the covariance matrix would be given by conventional uncertainties and Pearson's correlation parameter as desired; however, Pearson's correlation parameter depends not only on the method with which observables have been determined, but also on the conventional uncertainties such that this correlation parameter cannot be re-used in cases where only conventional uncertainties are available (see Sec.~\ref{sec:2-dim-corr}).

\begin{figure}
\begin{centering}
\includegraphics[width=9cm]{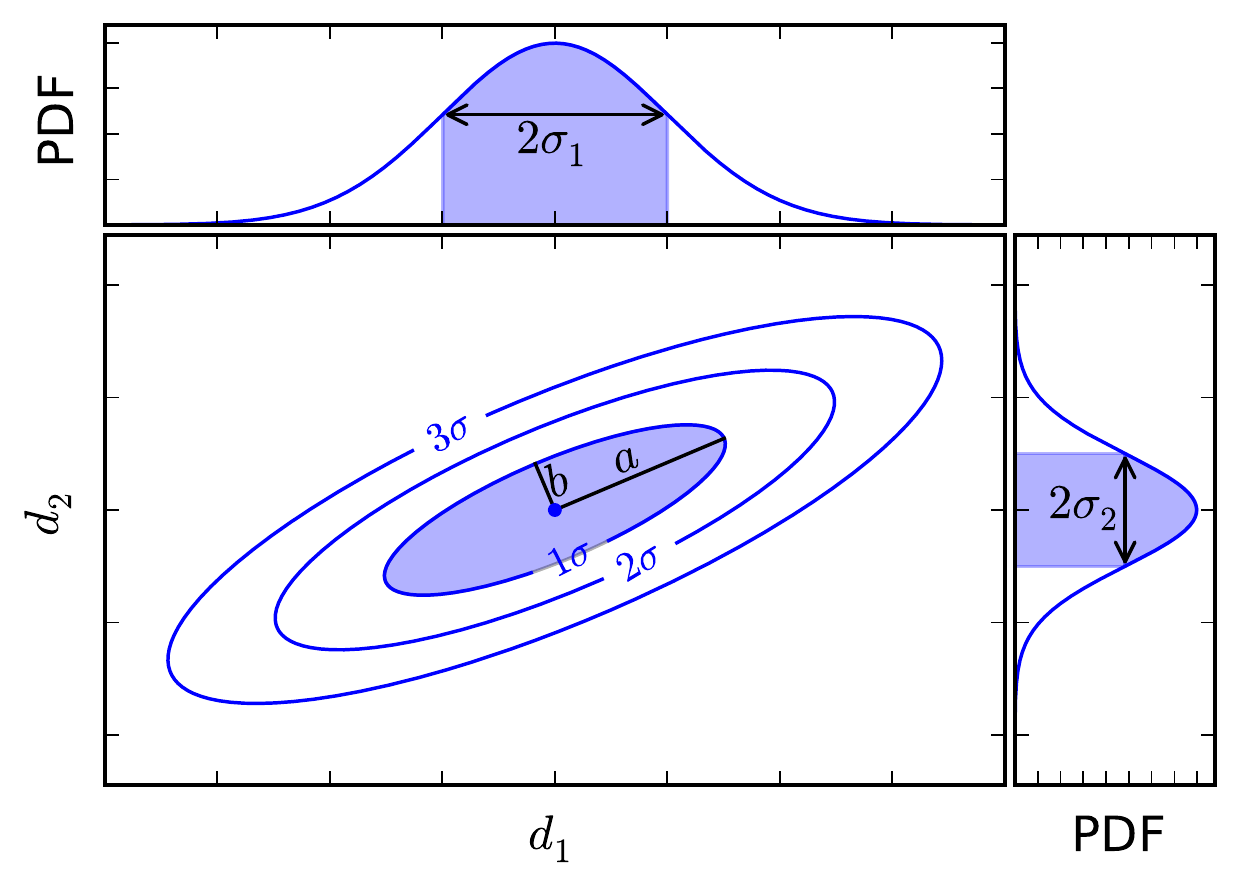}
\par\end{centering}
\caption{Schematic representation of the $1\sigma$, $2\sigma$, and $3\sigma$ contours of a Gaussian likelihood function of two correlated observables $d_1$ and $d_2$. The semi-major and -minor axes $a$ and $b$ of the $1\sigma$ contour are indicated. The marginalised probability density functions (PDFs) of the two observables are shown in the top and right panels. They are Gaussian functions with standard deviations $\sigma_1$ and $\sigma_2$ (the conventional error bars). In this example, $\sigma_1=2\sigma_2$.}
\label{fig:likelihood-cartoon}
\end{figure}

We therefore seek to derive an alternative representation of the covariance matrix for which correlation parameters can be re-used or derived analytically (see Sec.~\ref{sec:comp-likelihood-hrd}). To that end, we note that the two-dimensional version of the likelihood function defined in Eq.~\ref{eq:likelihood} has the shape of a rotated ellipse (Fig.~\ref{fig:likelihood-cartoon}). The semi-major and -minor axes $a$ and $b$ of this Gaussian ellipse define the orientation of a reference frame that is rotated with respect to the standard reference frame. Let $\mathcal{B}$ be the standard Euclidean basis and $\mathcal{T}$ the basis of the rotated reference frame (the basis $\mathcal{T}$ is given by the eigenvectors of $\cov$ and the eigenvalues are $a^2$ and $b^2$). In the rotated reference frame, the covariance matrix is diagonal and Eq.~(\ref{eq:likelihood}) can be written as a product of independent Gaussians whose variances are the eigenvalues of the covariance matrix.

There is a linear map $\rot:\;\mathcal{B} \rightarrow \mathcal{T}$ that transforms from basis $\mathcal{B}$ to $\mathcal{T}$ and that can be expressed as a rotation matrix. Using this transformation $\rot$ we can change the basis of the inverse of the covariance matrix to switch between representations relative to the basis $\mathcal{B}$ and $\mathcal{T}$,
\begin{equation}
\cov_\mathcal{B}^{-1} = \rot\, \cov_\mathcal{T}^{-1}\, \rot^\mathrm{T}, 
\label{eq:cov-transform}
\end{equation}
where $\cov_\mathcal{B}$ and $\cov_\mathcal{T}$ denote the covariance matrix with respect to basis $\mathcal{B}$ and $\mathcal{T}$, respectively. In most applications the conventional uncertainties $\vec{\sigma}$ of each observable are known but not the standard deviations of the likelihood in the rotated reference frame ($\sigma_1$ and $\sigma_2$ vs. $a$ and $b$; cf. Fig.~\ref{fig:likelihood-cartoon}). To compute one from the other, we derive relations between the correlation parameters, the conventional error bars, and the standard deviations of the rotated Gaussian likelihood by integrating the likelihood function in Eq.~(\ref{eq:likelihood}). In Sec.~\ref{sec:2-dim-corr} we derive the expressions for two correlated observables and in Sec.~\ref{sec:3-dim-corr} for three.

\subsubsection{Two correlated observables}\label{sec:2-dim-corr}
For two observables the rotation matrix $\rot$ can be written as
\begin{equation}
\rot =
\begin{pmatrix}
\cos \varphi & -\sin \varphi \\
\sin \varphi & \cos \varphi
\end{pmatrix}
\label{eq:R-2d}
\end{equation}
with $\varphi$ being the rotation angle by which the reference frame $\mathcal{T}$ is rotated with respect to $\mathcal{B}$; we call $\varphi$ the correlation parameter. Let $a$ and $b$ be the standard deviations of the rotated Gaussian such that the inverse of the covariance matrix with respect to basis $\mathcal{T}$ has the form
\begin{equation}
\cov_\mathcal{T}^{-1} =
\begin{pmatrix}
1/a^2 & 0 \\
0 & 1/b^2
\end{pmatrix}.
\label{eq:2d-inv-cov}
\end{equation}
The inverse of the covariance matrix with respect to basis $\mathcal{B}$ follows from a basis transformation (Eq.~\ref{eq:cov-transform}) using the rotation matrix $\rot$ (Eq.~\ref{eq:R-2d}). The likelihood function (Eq.~\ref{eq:likelihood}) then reads
\begin{equation}
p(\vec{d}|\vec{m})=\frac{1}{2\pi a b} \exp \left\{ -\frac{1}{2} \left[ \left(\frac{x'}{a}\right)^2 + \left(\frac{y'}{b}\right)^2 \right] \right\},
\label{eq:likelihood-2d}
\end{equation}
with
\begin{eqnarray}
x' &=& \left[ d_1 - d_1(\vec{m})\right] \cos \varphi + \left[ d_2 - d_2(\vec{m})\right] \sin \varphi, \nonumber \\
y' &=& \left[ d_1 - d_1(\vec{m})\right] \sin \varphi - \left[ d_2 - d_2(\vec{m})\right] \cos \varphi, \nonumber
\end{eqnarray}
and ($d_1$, $d_2$) and ($d_1(\vec{m})$, $d_2(\vec{m})$) are the components of $\vec{d}$ and $\vec{d}(\vec{m})$, respectively.

Integrating, \ie marginalising, the rotated Gaussian likelihood (Eq.~\ref{eq:likelihood-2d}) over $x\equiv d_1 - d_1(\vec{m})$ and $y\equiv d_2 - d_2(\vec{m})$ results again in Gaussian functions with standard deviations $\sigma_1 = \sqrt{a^2 \cos^2 \varphi + b^2 \sin^2 \varphi}$ and $\sigma_2 = \sqrt{b^2 \cos^2 \varphi + a^2 \sin^2 \varphi}$, respectively. These standard deviations are the conventional error bars that are typically reported in the literature, and solving the set of equations for $a$ and $b$ yields the desired relations for the standard deviations of the rotated Gaussian likelihood as a function of the conventional error bars of the observables and the correlation parameter,
\begin{eqnarray}
a^2 &=& \frac{-\sigma_1^2 \cos^2 \varphi + \sigma_2^2 \sin^2 \varphi}{\sin^4 \varphi - \cos^4 \varphi} \quad \text{and} \nonumber \\
b^2 &=& \frac{\sigma_1^2 \sin^2 \varphi - \sigma_2^2 \cos^2 \varphi}{\sin^4 \varphi - \cos^4 \varphi}.
\label{eq:sigmas-2d}
\end{eqnarray}
The above transformation from $(\sigma_1,\sigma_2)$ to $(a, b)$ only has unique solutions for $\varphi\neq\pm\pi/4$. In case of $\varphi=\pi/4$, $\sigma_1=\sigma_2\equiv\sigma$, and $\sigma^2=(a^2+b^2)/2$. As a consequence $a$ and $b$, \ie the covariance matrix, cannot be uniquely determined from the provided conventional uncertainties $\sigma_1$ and $\sigma_2$, and instead the full covariance matrix has to be provided in order to use the new likelihood model. The covariance matrix is symmetric and positive-definite, \ie its eigenvalues are real and positive: $a^2\geq 0$ and $b^2\geq 0$. For $a^2\rightarrow 0$ or $b^2\rightarrow 0$ the likelihood function describes a straight line in the $d_1$--$d_2$ plane.

The eccentricity, $e$, of the shape of the likelihood function is given by ($a>b$)
\begin{equation}
e^2 = 1-\frac{b^2}{a^2} = \frac{\sigma_1^2 - \sigma_2^2}{\sigma_1^2 \cos^2 \varphi - \sigma_2^2 \sin^2 \varphi},
\label{eq:eccentricity}
\end{equation}
showing that the likelihood function has the shape of a circle for $\sigma_1=\sigma_2$ ($\varphi\neq \pm \pi/4$) and that of an ellipse otherwise. For fixed $\sigma_1$ and $\sigma_2$, the orientation and eccentricity of the ellipse are determined by the correlation parameter $\varphi$.

The correlation parameter describes how two observables $x$ and $y$ co-vary, \ie by how much $y$ has to change when changing $x$ while maintaining the best possible fit/highest likelihood. Mathematically, the correlation parameter is therefore given by the derivative of $y$ with respect to $x$, 
\begin{equation}
\tan \varphi = \frac{\mathrm{d} y}{\mathrm{d} x}.
\label{eq:def-corr-param}
\end{equation}
For $\varphi << 1$ (as is the case in our examples below), Eq.~(\ref{eq:def-corr-param}) simplifies to $\varphi \approx \mathrm{d} y/\mathrm{d} x$.

Using our parametrisation, we find for the covariance of two observables $p$ and $q$,
\begin{equation}
\mathrm{Cov}(p,q) = \frac{1}{2} \left( \sigma_p^2 - \sigma_q^2 \right) \tan 2\varphi,
\label{eq:cov}
\end{equation}
and thus for Pearson's correlation parameter
\begin{equation}
\rho_{pq} = \frac{1}{2} \frac{\sigma_p^2 - \sigma_q^2}{\sigma_p \sigma_q} \tan 2\varphi.
\label{eq:pearson-corr-coeff}
\end{equation}
These relations show the advantage of our correlation parameter compared to other common representations of the covariance matrix, namely that $\varphi$ depends only on the method with which the observables $p$ and $q$ have been derived, which is not the case for the covariance and Pearson's correlation parameter.

\subsubsection{Three correlated observables}\label{sec:3-dim-corr}
In Sec.~\ref{sec:2-dim-corr} we derive a parametrisation of the likelihood function Eq.~(\ref{eq:likelihood}) for two correlated observables. This is already an important step because it allows us to consider correlations, for example in the HR diagram and the $T_\mathrm{eff}$--$\log g$ plane (Kiel diagram). However, there are often three correlated observables, for example if effective temperatures, surface gravities, and luminosities of stars are known simultaneously. Analogously to Sec.~\ref{sec:2-dim-corr}, we now derive the relations between the conventional error bars, the eigenvalues of $\cov$, and the correlation parameters for three correlated observables.

The shape of the likelihood function in Eq.~\ref{eq:likelihood} is now an ellipsoid whose orientation in space can be described by two angles, $\varphi$ and $\theta$. Let $\varphi$ be the rotation angle along the $z$-axis of the standard Euclidean basis $\mathcal{B}$ and $\theta$ the rotation angle along the rotated $y$-axis of $\mathcal{B}$. Furthermore, let $\mathcal{T}_1$ be the basis after rotating $\mathcal{B}$ along the $z$-axis by $\varphi$ and $\mathcal{T}_2$ the basis of the reference frame after both rotations. The overall linear map describing the two rotations is then
\begin{equation}
\vec{R}\equiv\vec{R}(\varphi,\theta)=\vec{R}_{y'}(\theta)\,\vec{R}_{z}(\varphi) = \vec{R}_{z}(\varphi)\,\vec{R}_{y}(\theta),
\label{eq:3-dim-rot}
\end{equation}
where $\vec{R}_{y'}(\theta)$ denotes a rotation by $\theta$ along the rotated $y$-axis of $\mathcal{B}$ (\ie the $y$-axis of $\mathcal{T}_1$), $\vec{R}_{z}(\varphi)$ a rotation by $\varphi$ along the $z$-axis of $\mathcal{B}$, and $\vec{R}_{y}(\theta)$ a rotation by $\theta$ along the $y$-axis of $\mathcal{T}_1$. In the second step in Eq.~(\ref{eq:3-dim-rot}), we have used that $\vec{R}_{y}(\theta)=\vec{R}_{z}^\mathrm{T}(\varphi)\,\vec{R}_{y'}(\theta)\,\vec{R}_{z}(\varphi)$. Overall this means that rotating first along the $z$-axis by $\varphi$ and then by $\theta$ along the rotated $y$-axis of $\mathcal{B}$ is equivalent to first rotating by $\theta$ along the $y$-axis and then by $\varphi$ along the $z$-axis of $\mathcal{B}$. In matrix form, Eq.~(\ref{eq:3-dim-rot}) reads
\begin{equation}
\vec{R} =
\begin{pmatrix}
\cos \varphi & -\sin \varphi & 0 \\
\sin \varphi & \cos \varphi & 0 \\
0 & 0 & 1
\end{pmatrix}
\begin{pmatrix}
\cos \theta & 0 & \sin \theta \\
0 & 1 & 0 \\
-\sin \theta & 0 & \cos \theta
\end{pmatrix}.
\label{eq:3d-rot-matrix-form}
\end{equation}

Using Eqs.~(\ref{eq:cov-transform}) and~(\ref{eq:3d-rot-matrix-form}), we compute the inverse of the symmetric covariance matrix (the covariance matrix with respect to $\mathcal{T}_2$ is diagonal and we assume the eigenvalues on the diagonal to be $a^2$, $b^2$, and $c^2$; cf. Eq.~\ref{eq:2d-inv-cov}) such that the likelihood function (Eq.~\ref{eq:likelihood}) is given by
\begin{equation}
p(\vec{d}|\vec{m})=\frac{1}{(2\pi)^{3/2} a b c} \exp \left\{ -\frac{1}{2} \left[ \left(\frac{x'}{a}\right)^2 + \left(\frac{y'}{b}\right)^2 + \left(\frac{z'}{c}\right)^2 \right] \right\}
\label{eq:likelihood-3d}
\end{equation}
with
\begin{align} 
\begin{split}
x' = {}& \left[ d_1 - d_1(\vec{m})\right] \cos \varphi \cos \theta + \left[ d_2 - d_2(\vec{m})\right] \cos \theta \sin \varphi \\ 
        & - \left[ d_3 - d_3(\vec{m})\right] \sin \theta,
\end{split} \nonumber \\
y' = {}& -\left[ d_1 - d_1(\vec{m})\right] \sin \varphi + \left[ d_2 - d_2(\vec{m})\right] \cos \varphi, \nonumber \\
\begin{split}
z' = {}& \left[ d_1 - d_1(\vec{m})\right] \cos \varphi \sin \theta + \left[ d_2 - d_2(\vec{m})\right] \sin \varphi \sin \theta \\
        & + \left[ d_3 - d_3(\vec{m})\right] \cos \theta .
\end{split} \nonumber
\end{align}
In order to compute $a$, $b$, and $c$ from the conventional uncertainties of each observable and the two rotation angles $\varphi$ and $\theta$, Eq.~(\ref{eq:likelihood-3d}) has to be integrated. To this end, we define $x=d_1 - d_1(\vec{m})$, $y=d_2 - d_2(\vec{m})$, and $z=d_3 - d_3(\vec{m})$ and rewrite Eq.~(\ref{eq:likelihood-3d}) as
\begin{align}
p(\vec{d}|\vec{m}) = {}& \frac{1}{(2\pi)^{3/2} a b c} \exp \left\{ -\frac{1}{2} \left[ \left( \frac{\alpha_1 x + \alpha_2 y + \alpha_3 z}{a} \right)^2 \right. \right. \nonumber \\
& \left. \left. + \left( \frac{\beta_1 x + \beta_2 y}{b} \right)^2 + \left( \frac{\gamma_1 x + \gamma_2 y + \gamma_3 z}{c} \right)^2 \right] \right\}
\label{eq:likelihood-3d-b}
\end{align}
with
\begin{eqnarray}
\alpha_1 &=& \cos \varphi \cos \theta, \nonumber \\
\alpha_2 &=& \cos \theta \sin \varphi, \nonumber \\
\alpha_3 &=& -\sin \theta, \nonumber \\
\beta_1 &=& -\sin \varphi, \nonumber \\
\beta_2 &=& \cos \varphi, \nonumber \\
\gamma_1 &=& \cos \varphi \sin \theta, \nonumber \\
\gamma_2 &=& \sin \varphi \sin \theta, \nonumber \\
\gamma_3 &=& \cos \theta.
\label{eq:new-definitions}
\end{eqnarray}
With these definitions we find
\begin{eqnarray}
L(x) &\equiv& \int_{-\infty}^{+\infty} \int_{-\infty}^{+\infty}\, p(\vec{d}|\vec{m}) \, \mathrm{d}y\,\mathrm{d}z = \frac{1}{\sqrt{2 \pi} \sigma_1} \exp \left[ -\frac{1}{2} \left( \frac{x}{\sigma_1} \right)^2\right], \nonumber \\
L(y) &\equiv& \int_{-\infty}^{+\infty} \int_{-\infty}^{+\infty}\, p(\vec{d}|\vec{m}) \, \mathrm{d}x\,\mathrm{d}z = \frac{1}{\sqrt{2 \pi} \sigma_2} \exp \left[ -\frac{1}{2} \left( \frac{y}{\sigma_2} \right)^2\right], \nonumber \\
L(z) &\equiv& \int_{-\infty}^{+\infty} \int_{-\infty}^{+\infty}\, p(\vec{d}|\vec{m}) \, \mathrm{d}x\,\mathrm{d}y = \frac{1}{\sqrt{2 \pi} \sigma_3} \exp \left[ -\frac{1}{2} \left( \frac{z}{\sigma_3} \right)^2\right], \nonumber
\end{eqnarray}
which are Gaussian functions with standard deviations
\begin{eqnarray}
\sigma_1^2 &=& a^2 \beta_2^2 \gamma_3^2 + b^2 (\alpha_3 \gamma_2 - \alpha_2 \gamma_3)^2 + c^2 \alpha_3^2 \beta_2^2, \nonumber \\
\sigma_2^2 &=& a^2 \beta_1^2 \gamma_3^2 + b^2 (\alpha_3 \gamma_1 - \alpha_1 \gamma_3)^2 + c^2 \alpha_3^2 \beta_1^2, \nonumber \\
\sigma_3^2&=&a^2 (\beta_1 \gamma_2 - \beta_2 \gamma_1)^2 + b^2 (\alpha_1 \gamma_2 - \alpha_2 \gamma_1)^2 + c^2 (\alpha_1 \beta_2 - \alpha_2 \beta_1)^2. \nonumber
\end{eqnarray}
Using the definitions from Eqs.~(\ref{eq:new-definitions}) and solving for the semi-major and -minor axes $a$, $b$, and $c$ of the ellipsoid, we arrive at the desired relation between the conventional uncertainties of each observable and the semi-major and -minor axes,
\begin{eqnarray}
a^2 &=& \frac{\cos^2\theta \left(\sigma_1^2 \cos^2\varphi - \sigma_2^2 \sin^2 \varphi \right) - \sigma_3^2 \sin^2 \theta \left( \cos^2\varphi - \sin^2 \varphi \right)}{\left( \cos^2 \varphi - \sin^2 \varphi \right) \left(\cos^2 \theta - \sin^2 \theta \right)}, \nonumber \\
b^2 &=& \frac{\sigma_1^2 \sin^2 \varphi - \sigma_2^2 \cos^2 \varphi}{\sin^4 \varphi - \cos^4 \varphi}, \nonumber \\
c^2 &=& \frac{\sigma_3^2 \cos^2 \theta \left(\cos^2 \varphi - \sin^2 \varphi \right) - \sin^2 \theta \left( \sigma_1^2 \cos^2\varphi - \sigma_2^2 \sin^2 \varphi \right)}{\left( \cos^2 \varphi - \sin^2 \varphi \right) \left( \cos^2 \theta - \sin^2 \theta \right)}. \nonumber
\label{eq:sigmas-3d}
\end{eqnarray}
As in the case of two correlated observables (Sec.~\ref{sec:2-dim-corr}), the full specification of our new likelihood model from conventional uncertainties is only possible if $\varphi \neq \pm \pi/4$ and $\theta \neq \pm \pi/4$. Furthermore, the combinations of $\sigma_1$, $\sigma_2$, $\sigma_3$, $\varphi$, and $\theta$ have to be such that $a^2>0$, $b^2>0$, and $c^2>0$ because $\cov$ is symmetric and positive-definite. 

We define the correlation parameters to describe correlations of two observables with respect to $\mathcal{B}$. With the definitions of $\varphi$ and $\theta$ as rotations along the $z$-axis and the rotated $y$-axis, $\varphi$ already describes the correlation between two observables in the $x\text{--}y$ plane, but $\theta$ does not describe the correlation in either the $x\text{--}z$ or $y\text{--}z$ plane of $\mathcal{B}$, but in the rotated $x\text{--}z$ plane of $\mathcal{B}$. Hence, if $\xi$ is the correlation parameter of two observables in the $x\text{--}z$ plane, $\theta$ and $\xi$ are related through
\begin{equation}
\tan \theta = \tan \xi \cos \varphi.
\label{eq:theta-3d-vs-2d}
\end{equation}

\subsection{Determination of correlation parameters}\label{sec:det-corr-param}

Whether observables co-vary or not depends solely on how they are determined. Observables can also be fully independent of each other, which is the case if it is possible to determine one parameter without needing to know the others or if observables are inferred from independent methods. On the contrary, it may be that the determination of parameters depends (strongly) on knowing the others, in which case the correlation between the observables may add valuable information when comparing them to stellar models.

Determining correlation parameters can be complex. In all probability, the easiest situation is if observables are related to each other via an analytical function that is used to derive some observables from others. In this case the correlation parameters follow from Eq.~(\ref{eq:def-corr-param}). We discuss such an example in Sec.~\ref{sec:comp-likelihood-hrd}.

If there is no analytical relation, the correlation parameter can be found by computing the covariance of those observables that will be matched against stellar models. To this end, the relevant parameter space has to be sampled such that a (multi-dimensional) probability distribution for the relevant observables can be computed. The correlation parameter can then be determined by fitting our new likelihood model for two correlated observables (Eq.~\ref{eq:likelihood-2d}) to the corresponding marginalised probability distribution. This method can always be applied, but may be computationally expensive. In Sec.~\ref{sec:comp-likelihood-kiel} we discuss such an example and also show how it is possible to determine the correlation parameter more easily.

\subsection{Correlations in the Hertzsprung--Russell diagram}\label{sec:comp-likelihood-hrd}

\begin{figure}
\begin{centering}
\includegraphics[width=9cm]{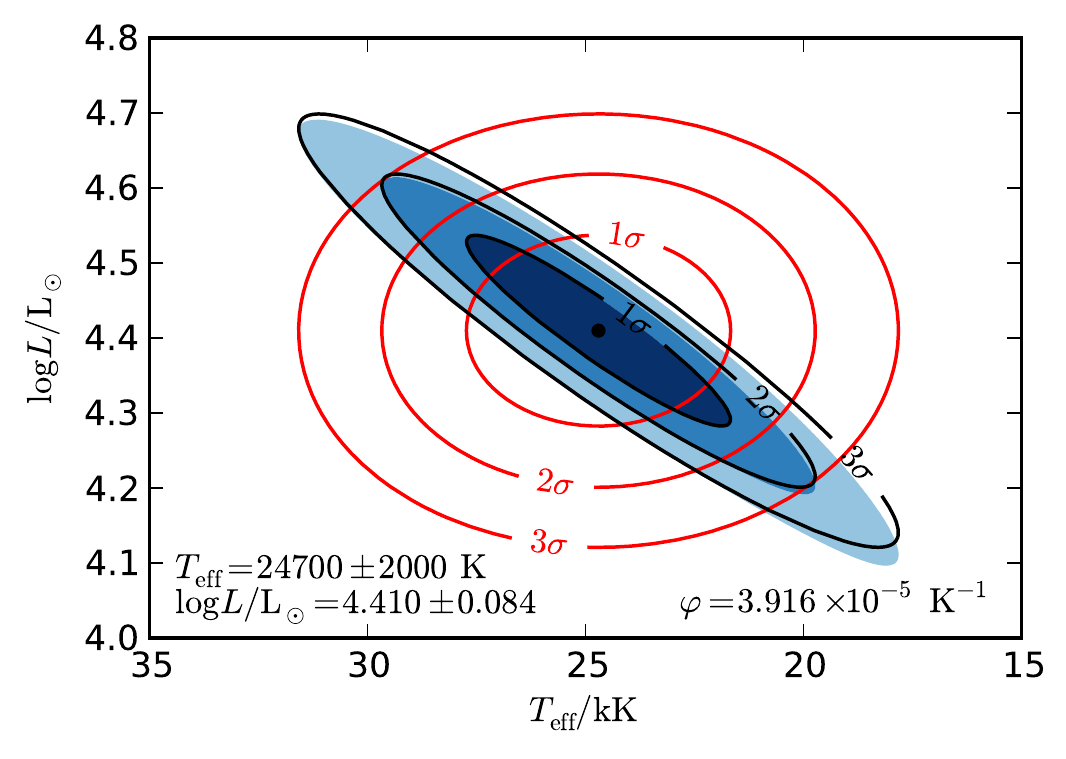}
\par\end{centering}
\caption{Comparison of the reference likelihood function with two approximations that neglect and account for correlations. The $1\sigma$, $2\sigma$, and $3\sigma$ areas of the reference likelihood function in the HR diagram are shown by the shaded areas. The red contours show the same confidence levels of a likelihood model that neglects the correlation in luminosity and effective temperature, and the black contours show our new likelihood model for a correlation parameter of $\varphi=3.916\times10^{-5}\,\mathrm{K}^{-1}$. The marginalised observables are $T_\mathrm{eff}=24700\pm2000\,\mathrm{K}$ and $\log L/\lsun=4.410\pm0.084$ for all three likelihood models.}
\label{fig:logl-teff-corr-map}
\end{figure}

Our new likelihood model can be applied to any kind of star in any situation as long as the correlations are known. In the following, we turn to massive, main-sequence stars because such models are already available in \bonnsai. We consider the situation that the effective temperature of a star is known, for example from fitting its spectral energy distribution or using a spectral type calibration, the apparent V-band magnitude from photometric observations, and the distance from parallax measurements. The luminosity of the star can then be computed using the bolometric correction calibration of \citet{1996ApJ...469..355F}. We choose the bolometric correction of \citet{1996ApJ...469..355F} for demonstration purposes because it is valid over a wide range of stellar temperatures, from M- to O-type stars. For this example, we assume that the effective temperature is $T_\mathrm{eff}=24700\pm2000\,\mathrm{K}$, the distance to the star $d=151\pm5\,\mathrm{pc}$, the apparent V-band magnitude $m_\mathrm{V}=1.97\pm0.01$, and that there is only negligible extinction (these observables, except for the large uncertainty in effective temperature, are characteristic of the magnetic B-type star HD~44743; \citealt{2015A&A...574A..20F}). The luminosity of the star is then given by
\begin{equation}
\log L/\lsun = 0.4\left( M_{\mathrm{V},\odot} - m_\mathrm{V} + 5\log \frac{d}{\mathrm{pc}} - 5 - BC_\mathrm{V}(T_\mathrm{eff}) \right),
\label{eq:bol-lum}
\end{equation}
where $BC_\mathrm{V}(T_\mathrm{eff})$ is the bolometric correction and $M_{\mathrm{V},\odot}$ the absolute V-band magnitude of the Sun. This equation shows that the logarithmic luminosity co-varies with apparent magnitude, distance, and effective temperature. On a more fundamental level, the reason why luminosity and effective temperature co-vary is that both quantities are given by the total flux of stars such that the Stefan--Boltzmann law holds, $L=4\pi R^2 \sigma T_\mathrm{eff}^4$ with $R$ being the stellar radius and $\sigma$ the Stefan--Boltzmann constant. We now consider the case where the luminosity and effective temperature are matched against stellar models to infer fundamental stellar parameters for this star. We are therefore only interested in the covariance of luminosity and effective temperature, and show in Fig.~\ref{fig:logl-teff-corr-map} the likelihood function computed from Eq.~\ref{eq:bol-lum} using the above stellar parameters and assuming Gaussian uncertainties; the marginalised luminosity is then $\log L/\lsun=4.410\pm0.084$. This likelihood model makes the fewest number of assumptions on the data and is therefore the most accurate model discussed here. In order to easily distinguish between the different likelihood models, we call it the reference likelihood model from here on. We further show the old Gaussian likelihood model that neglects correlations for $T_\mathrm{eff}=24700\pm2000\,\mathrm{K}$ and $\log L/\lsun=4.410\pm0.084$. The two likelihood functions differ significantly. 

\begin{figure*}[ht]
\begin{centering}
\includegraphics[width=9cm]{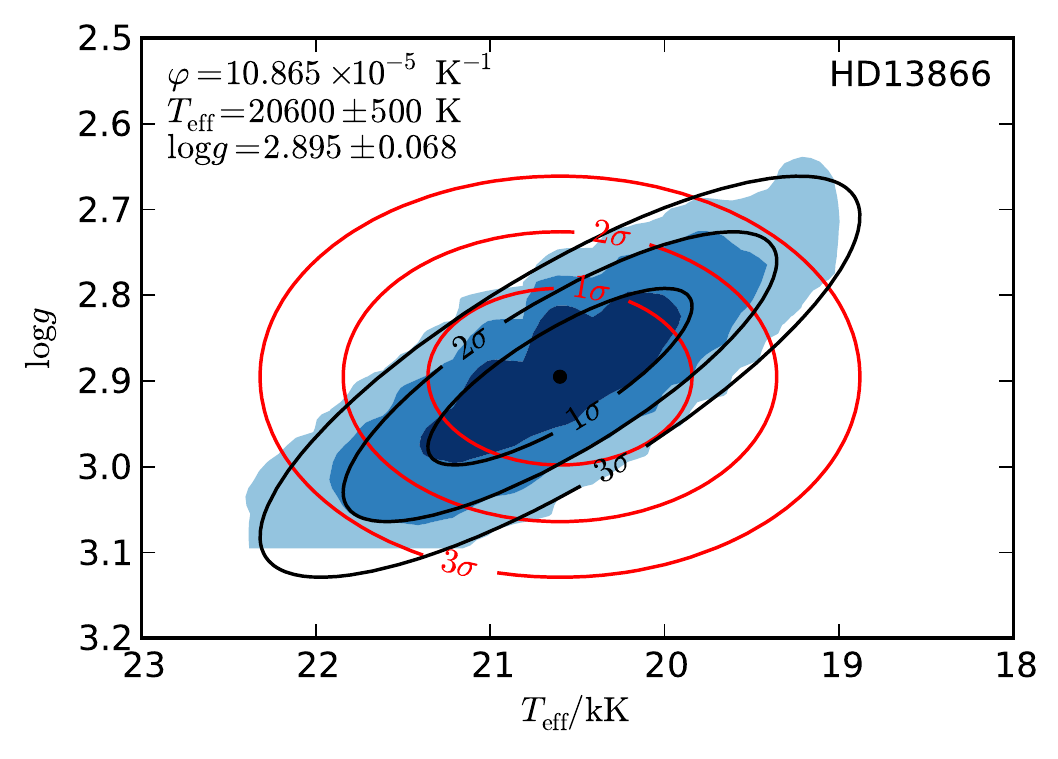}
\includegraphics[width=9cm]{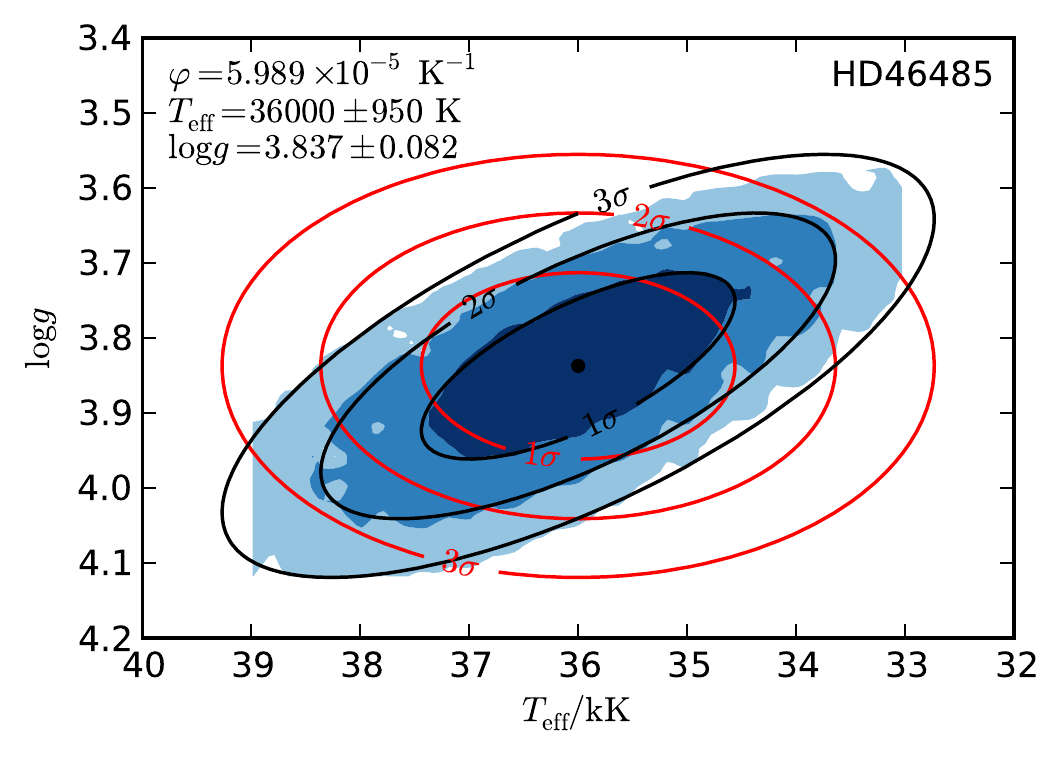}
\par\end{centering}
\caption{As in Fig.~\ref{fig:logl-teff-corr-map} but for (a) HD~13866 and (b) HD~46485 in the Kiel diagram. The probability maps have been derived by fitting \fastwind atmosphere models to observed spectra.}
\label{fig:logg-teff-corr-map}
\end{figure*}

Our new likelihood model is parametrised by the same luminosity and effective temperature as the old model with the addition of a correlation parameter $\varphi$ that describes how the luminosity co-varies with effective temperature. Here, $\varphi << 1$, and applying Eq.~(\ref{eq:def-corr-param}) we find
\begin{equation}
\varphi(T_\mathrm{eff}) = \frac{\mathrm{d}\log L/\lsun}{\mathrm{d}T_\mathrm{eff}} = -0.4\frac{\mathrm{d}BC}{\mathrm{d}T_\mathrm{eff}}.
\end{equation}
Using the analytic fit of the bolometric correction as a function of effective temperature from \citet{1996ApJ...469..355F}, \ie taking the derivative of the fit of the bolometric correction with respect to effective temperature and evaluating the derivative at $T_\mathrm{eff}=24700\,\mathrm{K}$, we find $\varphi=3.916\times10^{-5}\,\mathrm{K}^{-1}$. It is evident from Fig.~\ref{fig:logg-teff-corr-map} that the new likelihood model represents the reference likelihood well and much better than the old model.

Because of the different shape and orientation between the new likelihood model and the old, certain combinations of effective temperature and luminosity can be excluded with more than $3\sigma$ confidence, which would be considered possible within $1\sigma$ when neglecting the correlations. Given that different combinations of effective temperature and luminosity correspond to different combinations of the fundamental stellar parameters initial mass, age, and initial rotational velocity, this will influence the inference of these parameters (cf. Fig.~\ref{fig:implications-examples} and Sec.~\ref{sec:results} where we show that the most likely stellar parameters and the inferred uncertainties can change significantly).

Although the new likelihood model approximates the reference likelihood far better than a likelihood model that neglects correlations, the new approximation is not perfect. The correlation parameter varies with effective temperature leading to small deviations from the reference likelihood visible only in the $2\text{--}3\sigma$ regions. Halving the large uncertainty in effective temperature makes the deviations disappear nearly completely. We therefore conclude that the new likelihood model is an adequate approximation in this case.

\subsection{Correlations in the Kiel diagram}\label{sec:comp-likelihood-kiel}

The second example of correlations concerns effective temperatures and surface gravities that are deduced from fitting synthetic spectra of atmosphere models to observed ones. The correlation between surface gravity and effective temperature from fitting the spectra of stars depends on the details of the fitting process and the applied methods (see below). For example, in OB stars the correlation is such that hotter temperatures require larger gravities to fit spectra similarly well, whereas the opposite is true in cooler stars. In Wolf--Rayet stars, the photosphere is formed in the wind and the spectral lines are therefore not sensitive to the surface gravity. Hence, effective temperature and surface gravity are uncorrelated in these stars (in fact, the surface gravity remains mostly unconstrained). These examples show that the correlations of observables depend only on how observables are determined (cf. Sec.~\ref{sec:det-corr-param}).

In Fig.~\ref{fig:logg-teff-corr-map} we show how effective temperature and surface gravity co-vary when modelling the spectra of HD13866 and HD46485, observed within the IACOB project \citep{2011BSRSL..80..514S,2015hsa8.conf..576S}, with the stellar atmospheric code \fastwind \citep{1997A&A...323..488S,2005A&A...435..669P}. The probability distributions in Fig.~\ref{fig:logg-teff-corr-map} have been computed in two steps. First, the main optical transitions in the observed spectra were compared to a pre-computed grid of \fastwind stellar atmosphere models presented in \citet{2012A&A...542A..79C} to find the best-fitting spectroscopic stellar parameters \citep[for further details on the technique and the lines used, see][]{2012A&A...542A..79C}. Second, sub-grids of atmosphere models have been computed around the best-fitting effective temperature and surface gravity to properly resolve the probability distributions such that the correlation parameter can be reliably determined. The finer sub-grids have a resolution of $\Delta T_\mathrm{eff}=250\,\mathrm{K}$ and $\Delta \log g=0.025\,\mathrm{dex}$ and the probability map is a bit patchy because of this finite grid in atmosphere models. From the probability maps we find correlation parameters of $10.865\times 10^{-5}\,\mathrm{K}^{-1}$ and $5.989\times 10^{-5}\,\mathrm{K}^{-1}$ for HD~13866 and HD~46485, respectively. In this example, the correlation is stronger in the cooler OB star.

As in Sec.~\ref{sec:comp-likelihood-hrd} for a star in the HR diagram, it is evident that correlations can significantly change the shape and orientation of the likelihood function. The new likelihood model nicely matches the probability maps and properly reproduces the different confidence regions. A comparison with the old likelihood model reveals the need for taking correlations properly into account to avoid biases when determining fundamental stellar parameters such as mass and age.

Determining the correlation parameter is not trivial. As discussed in Sec.~\ref{sec:det-corr-param}, a viable method to obtain the correlation parameter is to fit our likelihood model to detailed probability maps such as those in Fig.~\ref{fig:logg-teff-corr-map}; this procedure can be computationally expensive. An alternative, computationally less expensive method works as follows: First, the best-fitting spectroscopic parameters are determined by varying those parameters that significantly influence the diagnostic lines. Second, the best-fitting effective temperature is changed by $\delta T_\mathrm{eff}$ and $\log g$ is varied by $\delta \log g$ until the deviation of the synthetic spectrum from the observed spectrum is minimised again. During this step, all other parameters can be kept constant at the best-fitting values and the correlation parameter follows from $\varphi=\delta \log g/\delta T_\mathrm{eff}$ because $\varphi << 1$ (Eq.~\ref{eq:def-corr-param}). It is advisable to repeat the steps for several values of $\delta T_\mathrm{eff}$ to obtain a more robust correlation parameter. Both procedures can of course be applied to any stellar parameter and not only effective temperature and surface gravity.

The absolute value of the correlation parameter of effective temperature and surface gravity depends on a variety of factors; it depends on which atmospheric lines are used in the analysis (\eg hydrogen--helium lines or hydrogen--helium--silicon lines) and which weight is given to the diagnostic lines. For example, the surface gravity is strongly constrained by the wings of the Balmer lines and some fitting methods require always fitting the Balmer wings well while allowing for more freedom in other lines. The Balmer wings and thus the derived surface gravity further depend on the signal-to-noise ratio (S/N), \ie the correlation parameter will also be a function of S/N. Finally, the correlation parameter depends on the atmospheric parameters varied in the fitting process because all parameters have the potential to change the shape and strengths of the main-diagnostic lines. For example, allowing for variations in the helium abundances obviously influences the helium lines and thus the correlation parameter; the wind-Q parameter also influences some helium lines \citep[\eg][]{1996A&A...305..171P,2000ARA&A..38..613K}.

\subsection{Normalised correlation parameter}\label{sec:normalised-correlation-parameter}

The deviation of the new likelihood model from a likelihood model that neglects correlations depends on the absolute value of the correlation parameter and it also depends on the conventional error bars. If a quantity $y$ co-varies strongly with $x$ but the conventional uncertainty of $x$ is much smaller than that of $y$, the new likelihood model deviates only slightly from a likelihood that neglects correlations. Similarly, the deviation of the likelihoods can be significant if $y$ co-varies only weakly with $x$ but the uncertainty of $x$ is much larger than that of $y$. The importance of the correlations therefore depends on the interplay between the correlation parameter and the error bars (cf. Eq.~\ref{eq:eccentricity}).

In order to judge the importance of correlations, it is useful to define a normalised correlation parameter, which is the correlation parameter introduced in Sec.~\ref{sec:2-dim-corr} divided by the ratio of the corresponding $1\sigma$ uncertainties:
\begin{equation}
\hat{\varphi} = \varphi / \frac{\sigma_2}{\sigma_1} = \frac{\mathrm{d} y}{\mathrm{d} x} / \frac{\sigma_2}{\sigma_1}.
\label{eq:normalised-correlation-parameter}
\end{equation}
This parameter is a measure of the importance of correlations with $\hat{\varphi}\rightarrow 0$ indicating no influence and $\hat{\varphi}\rightarrow 1$ maximum influence on the shape of the likelihood (in the former case, the likelihood function equals a likelihood that neglects correlations, whereas it converges to a straight line in the latter). 

A geometric interpretation of the normalised correlation parameter may be obtained by considering $\varphi << 1$. This applies for correlations between effective temperature and logarithmic luminosity and between effective temperature and logarithmic surface gravity discussed in Secs.~\ref{sec:comp-likelihood-hrd} and~\ref{sec:comp-likelihood-kiel}, respectively. The ratio of the $1\sigma$ areas of the likelihood function accounting for and neglecting correlations is then
\begin{equation}
\frac{\pi a b}{\pi \sigma_1 \sigma_2} \approx \sqrt{1-\left(1+\sigma_2^4/\sigma_1^4 \right)\hat{\varphi}^2} \approx \sqrt{1-\hat{\varphi}^2}.
\label{eq:geo-interp-norm-corr-param}
\end{equation}
The last approximation holds if $\sigma_2^4/\sigma_1^4<<1$, which is the case for the mentioned examples. In these cases, the normalised correlation parameter relates directly to the relative change in the $1\sigma$ area of the likelihood function. This interpretation also allows us to quantify the importance of correlations for the likelihood function: a larger than 5\% (10\%) reduction of the $1\sigma$ area corresponds to $\hat{\varphi} \gtrsim 0.31$ ($\hat{\varphi} \gtrsim 0.43$).

For the example of star HD~44743 in the HR diagram (Sec.~\ref{sec:comp-likelihood-hrd}), the normalised correlation parameter is $\hat{\varphi}=0.93$, indicating a strong correlation that has large consequences for the likelihood model. Indeed, the area of the $1\sigma$ region is, in accordance with Eq.~(\ref{eq:geo-interp-norm-corr-param}), reduced by about 63\% when accounting for correlations (Fig.~\ref{fig:logl-teff-corr-map}). In the Kiel diagram, the normalised correlation parameters are $\hat{\varphi}=0.8$ and $\hat{\varphi}=0.7$ for HD~13866 and HD~46485, respectively (Sec.~\ref{sec:comp-likelihood-kiel}), leading to a reduction of about 29\% and 40\% of the $1\sigma$ area when accounting for correlations (Fig.~\ref{fig:logg-teff-corr-map}).

The absolute correlation parameters of HD~13866 and HD~46485 in the Kiel diagram are larger than that of HD~44743 in the HR diagram (Secs.~\ref{sec:comp-likelihood-hrd} and~\ref{sec:comp-likelihood-kiel}). However, the shape of the likelihood of HD~44743 is more elongated (Fig.~\ref{fig:logl-teff-corr-map}), \ie the correlation is more important, than for HD~13866 and HD~46485 (Fig.~\ref{fig:logg-teff-corr-map}). This shows the interplay between the correlation parameter and the $1\sigma$ uncertainties, and highlights the value of the normalised correlation parameter in providing a degree for the strength of correlations.

\section{Results}\label{sec:results}

As shown in Secs.~\ref{sec:comp-likelihood-hrd} and~\ref{sec:comp-likelihood-kiel}, likelihood models that neglect and take correlations into account can cover significantly different regions of the parameter space and will therefore affect the determination of fundamental stellar parameters such as mass and age. To this end we investigate changes in the inferred initial masses and ages for a sample of nearly 500 mock stars distributed all over the HR and Kiel diagrams. The mock stars are taken from the Milky Way stellar models of \citet{2011A&A...530A.115B} and initial masses and ages are distributed in such a way to achieve a good coverage of the HR and Kiel diagrams. We choose $1\sigma$ uncertainties of $1000\,\mathrm{K}$, $0.1\,\mathrm{dex}$, and $0.1\,\mathrm{dex}$ in effective temperature, logarithmic luminosity, and logarithmic surface gravity, respectively. These uncertainties are characteristic values for spectroscopically derived stellar parameters. The correlation parameter of effective temperature and luminosity in Sec.~\ref{sec:comp-likelihood-hrd} is about $0.4\times 10^{-4}\,\mathrm{K}^{-1}$, resulting in a normalised correlation parameter of about $0.4$ for our choice of the $1\sigma$ uncertainties. We choose a normalised correlation parameter of $0.8$ for effective temperature and surface gravity reminiscent of those found for HD~13866 and HD~46485 in Sec.~\ref{sec:comp-likelihood-kiel}. We thus introduce one case of modest and one case of notable correlation (reductions in the $1\sigma$ area of about 8\% and 40\%, respectively), and therefore expect that the most likely values and precisions are more strongly affected for stars with known effective temperatures and gravities than for stars with known effective temperatures and luminosities. A different choice of error bars (\eg smaller uncertainties on luminosities) may lead to the opposite situation. We therefore caution that the quantitative results presented here are only valid for our particular choice of correlations \emph{and} error bars. 

In our analysis, we use a Salpeter initial mass function \citep{1955ApJ...121..161S} as initial mass prior distribution, a uniform age prior distribution, and a Gaussian initial rotational velocity prior distribution with a mean of $100\,\mathrm{km}\,\mathrm{s}^{-1}$ and full width at half maximum of $250\,\mathrm{km}\,\mathrm{s}^{-1}$ \citep[the initial rotational velocity prior distribution is thus reminiscent of the observed distribution of rotational velocities of B-type stars in the Milky Way,][]{2008A&A...479..541H}. \bonnsai samples all those Milky Way stellar models from \citet{2011A&A...530A.115B} from a pre-computed database that are within $5\sigma$ of the observations. Given that the mock stars are taken from the same stellar models and the chosen size of the observational error bars, the posterior-predictive check (goodness-of-fit test) and resolution test are passed in each case.

We first discuss the qualitative changes in inferred masses and ages because of correlations (Sec.~\ref{sec:qualitative-influence}) before presenting quantitative results for the precisions (Sec.~\ref{sec:precisions}) and most likely (Sec.~\ref{sec:most-likely-parameters}) initial masses and ages. Also, the choice of prior distributions influences the inference of mass and age, and we examine this in Appendix~\ref{sec:influence-priors}.

\subsection{Qualitative discussion of the influence of correlations on inferred stellar parameters}\label{sec:qualitative-influence}

\begin{figure*}
\begin{centering}
\includegraphics[width=18cm]{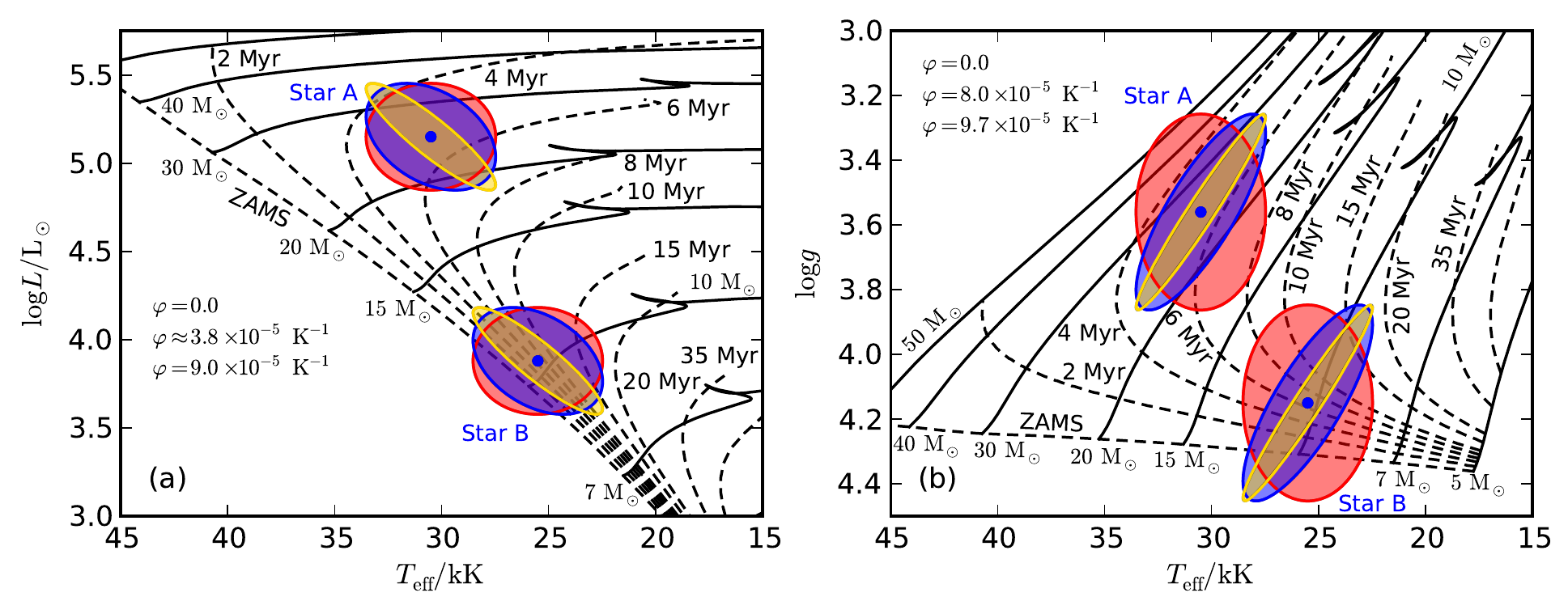}\par\end{centering}
\caption{Mock stars, Star~A and~B, in the HR diagram, panel (a), and Kiel diagram, panel (b). The error ellipses indicate $1\sigma$ confidence regions and we increased our standard error bars by a factor of 2 for illustration purposes. The red ellipses are for the case of uncorrelated observables and the blue ellipses for correlations as described in Secs.~\ref{sec:comp-likelihood-hrd} and~\ref{sec:comp-likelihood-kiel}, and applied in Secs.~\ref{sec:precisions} and~\ref{sec:most-likely-parameters}. For illustration purposes, the golden ellipses show very strong correlations. The solid lines are the non-rotating, Milky Way metallicity stellar tracks of \citet{2011A&A...530A.115B} for masses ranging from $5$ to $50\,\msun$, and the dashed lines are the corresponding isochrones from $0$ to $50\,\mathrm{Myr}$.}
\label{fig:implications-examples}
\end{figure*}

In order to illustrate the expected changes in inferred stellar parameters, we show the position of two of the approximately 500 mock stars, Star~A and~B, in the HR and Kiel diagram (Fig.~\ref{fig:implications-examples}). In Fig.~\ref{fig:implications-examples} we show the $1\sigma$ contours for (i) no correlations, (ii) the correlations used in the quantitative analysis in Secs.~\ref{sec:precisions} and~\ref{sec:most-likely-parameters} (see above), and (iii) very strong correlations. The error ellipses are not rotated for non-correlations and turn into straight lines for maximum correlations. 

Because of the correlations, the error ellipses change their orientation relative to the stellar tracks and isochrones. If an error ellipse is parallel to the stellar tracks (isochrones), the $1\sigma$ region covers fewer different masses (ages) than the error ellipse in the case of uncorrelated observables, meaning that the precision with which the mass (age) of a star is determined increases. The precision can also decrease if the error ellipse happens to be perpendicular to the stellar tracks (isochrones) such that a wider range of masses (ages) is covered. For Star~A, the stellar tracks and isochrones are highly inclined with respect to the semi-major axis of the error ellipse in the HR diagram and about parallel in the Kiel diagram. Consequently, mass and age can be determined to a lower (higher) precision from the HR (Kiel) diagram when taking correlations into account. For star~B in the HR diagram the age can be determined with a higher precision, while the mass is less precisely known when taking correlations into account. The situation is opposite in the Kiel diagram: the mass can now be determined to a higher precision, while age only to a lower precision.

Not only does the precision of inferred stellar parameters change, but also their most likely values. The most likely model takes the model density of stars because of different stellar lifetimes in various regions of the parameter space and prior knowledge such as the initial mass function into account. The most likely values of inferred stellar parameters are unaffected by correlations only if the model density is homogeneous and uniform
prior distributions are applied. However, the model density is not homogeneous, but increases towards less massive and younger stars, and initial-mass prior distributions usually favour lower mass stars. In the HR diagram the model density is therefore highest towards hotter effective temperatures and less luminous stars, and in the Kiel diagram towards cooler effective temperatures and higher gravities. The correlations discussed in Secs.~\ref{sec:comp-likelihood-hrd} and~\ref{sec:comp-likelihood-kiel} are exactly the opposite: hotter temperatures require brighter luminosities in the HR diagram and larger gravities in the Kiel diagram. Hence, the correlations work against the tendency that the most likely initial mass and age is found towards the highest model densities. Neglecting correlations can therefore bias inferred stellar parameters. 

We want to stress that no matter whether it is the precisions or the inferred most likely values that change when taking correlations into account, the reliability and robustness of the inferred fundamental stellar parameters and their uncertainties always improve. This may be counterintuitive, especially when parameters can only be determined to a lower precision when accounting for correlations, but it simply means that inferred uncertainties are underestimated otherwise.

\subsection{Precision of inferred initial masses and ages}\label{sec:precisions}

\subsubsection{Stars in the HR diagram}\label{sec:impact-hr-diagram}

\begin{figure*}
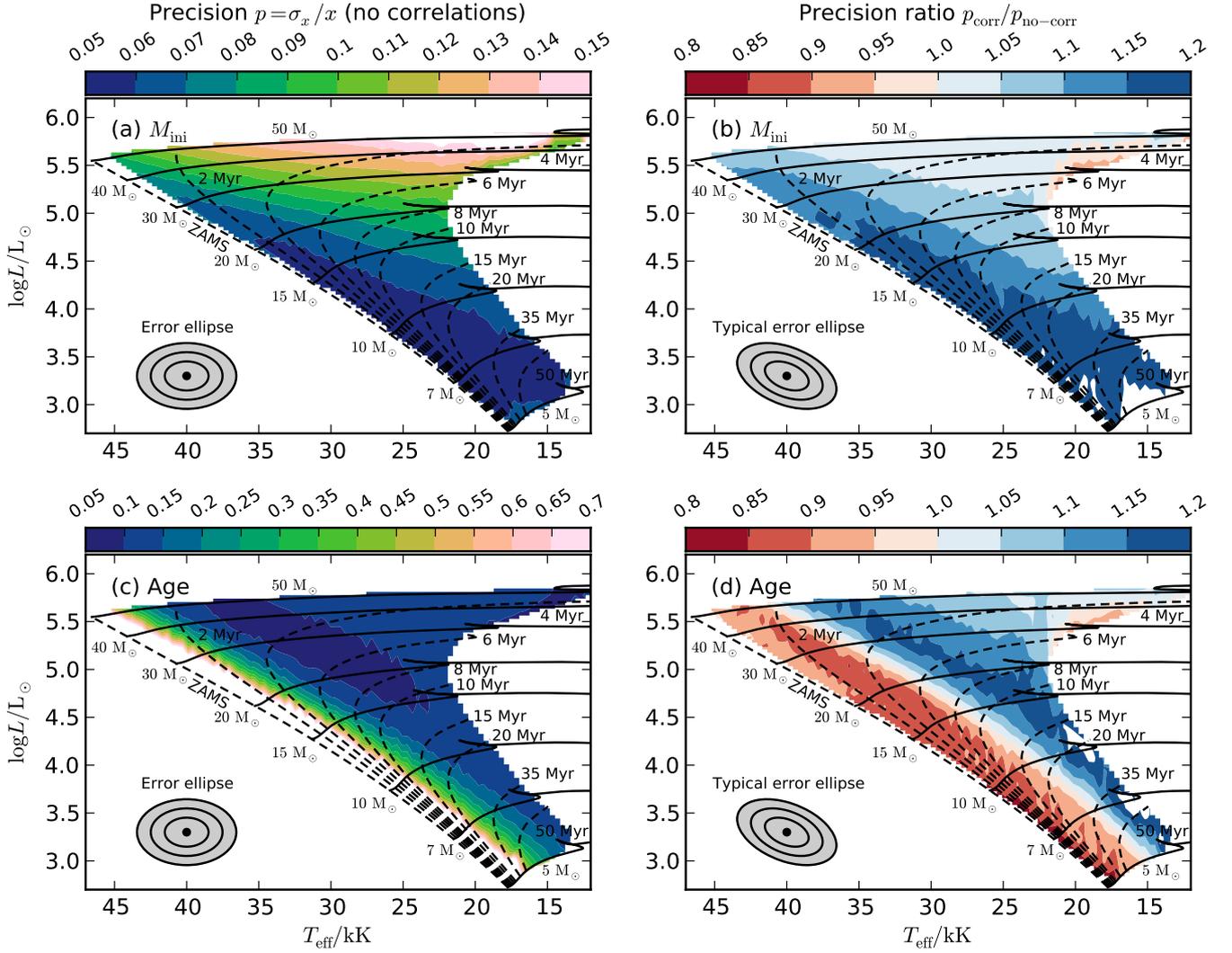

\begin{centering}
\includegraphics[width=18cm]{{{prec-ratio-map-tl_dTeff-1000_dlogL-0.10_dlogg-0.10-prec}}}
\par\end{centering}
\caption{Precisions $p=\sigma_x/x$ (left panels) and precision ratios $p_\mathrm{corr}/p_\mathrm{no-corr}$ (right panels) of determined initial masses (upper panels) and ages (lower panels) of our mock stars in the HR diagram (here, $\sigma_x$ and $x$ are the $1\sigma$ error bars and most likely values of initial mass and age, respectively, and $p_\mathrm{corr}$ and $p_\mathrm{no-corr}$ denote the precisions when considering and neglecting correlations, respectively). The colour-coding in the left panels (a) and (c) show the precision neglecting correlations and the colours in the right panels (b) and (d) the ratio of the precisions taking correlations into account. The $1\sigma$ uncertainties of the mock stars are $1000\,\mathrm{K}$ in effective temperature and $0.1\,\mathrm{dex}$ in luminosity, and the resulting $1\sigma$, $2\sigma$, and $3\sigma$ contours of the likelihood function are shown in the lower left corners (a correlation parameter of $4\times 10^{-5}\,\mathrm{K}^{-1}$ is used for illustration purposes in panels b and d). The plotted stellar tracks and isochrones are Milky Way metallicity, non-rotating models from \citet{2011A&A...530A.115B}.}
\label{fig:prec-hrd}
\end{figure*}

We explore the precision, $p$, with which the initial mass and age of our mock stars can be determined when neglecting correlations and then compare it to the case when taking correlations into account. The precision is defined as the ratio of the $1\sigma$ uncertainty and the mode (most likely) value of the posterior probability distributions; it is shown in the left panels (a) and (c) of Fig.~\ref{fig:prec-hrd}. In the right panels (b) and (d) of the same figure, we show the ratio of precisions taking correlations into account and neglecting them, $r=p_\text{corr}/p_\text{no corr}$. A ratio of $r>1$ indicates a lowering and a ratio of $r<1$ an improvement in the precisions when taking correlations into account. The ratio is defined such that the product of the ratios in panels (b) and (d), and the precisions in panels (a) and (c) directly give the precisions when taking correlations into account.

Neglecting correlations, the initial mass and age can be determined up to a precision of about 5\% for the given uncertainties in effective temperature and luminosity. The precision scales roughly linearly with the error bars, \ie halving the uncertainties of effective temperature and luminosity means that masses and ages can be determined two times more precisely. Masses and ages can be determined to the highest precision wherever the spacing between stellar tracks and isochrones is largest, giving rise to the gradients in Figs.~\ref{fig:prec-hrd}a and~\ref{fig:prec-hrd}c. Formally, the age precision diverges for stars approaching the zero age main sequence (ZAMS). We show precisions down to 70\%, resulting in uncoloured areas close to the ZAMS.

As illustrated in Fig.~\ref{fig:implications-examples}, taking correlations into account changes the precision with which mass and age can be determined depending on the relative orientation of the likelihood function to the stellar tracks and isochrones. The precision in mass always decreases because the error ellipses are highly inclined with respect to the stellar tracks such that the rotated error ellipses cover a wider range of masses; the decrease can be up to 20\% (Figs.~\ref{fig:prec-hrd}b and~\ref{fig:prec-hrd}d). The precision in age increases for stars close to the ZAMS because the orientation of the error ellipse is almost parallel to the isochrones and decreases once the isochrones bend over and become more parallel to the stellar tracks. The bending of the isochrones and the change in the precisions are nicely illustrated in Fig.~\ref{fig:prec-hrd}d and amount to up to about $\pm 20\%$.

\subsubsection{Stars in the Kiel diagram}\label{sec:impact-kiel-diagram}

\begin{figure*}
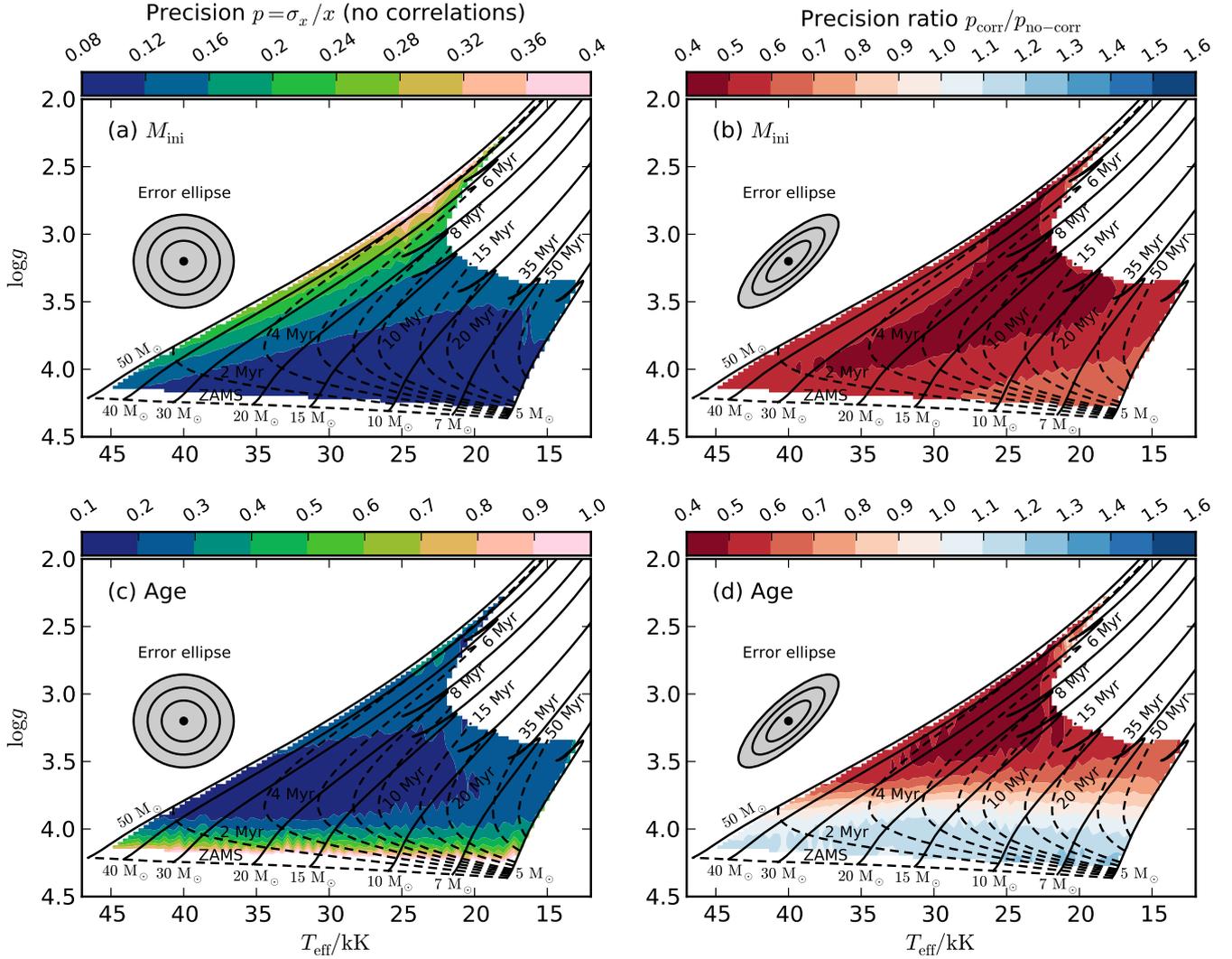

\begin{centering}
\includegraphics[width=18cm]{{{prec-ratio-map-tg_dTeff-1000_dlogL-0.10_dlogg-0.10-prec}}}
\par\end{centering}
\caption{Same as Fig.~\ref{fig:prec-hrd} but for stars in the Kiel diagram. The $1\sigma$ uncertainties are $1000\,\mathrm{K}$ in effective temperature and $0.1\,\mathrm{dex}$ in logarithmic surface gravity, and a correlation parameter of $0.8\times 10^{-4}\,\mathrm{K}^{-1}$ is assumed for all stars (cf. Sec.~\ref{sec:comp-likelihood-kiel}). The error ellipses show $1\sigma$, $2\sigma$, and $3\sigma$ uncertainty contours.}
\label{fig:prec-kiel}
\end{figure*}

Analogously to Sec.~\ref{sec:impact-hr-diagram}, we study the precision with which masses and ages can be determined from the position of stars in the Kiel diagram. Neglecting correlations, masses, and ages can be determined with a precision of up to about 10\% (Fig.~\ref{fig:prec-kiel}). As discussed in Sec.~\ref{sec:impact-hr-diagram}, the precision in mass and age is the best for stars in those areas of the Kiel diagram where the spacing between stellar tracks and isochrones is the largest. 

Unlike stars in the HR diagram, the precision with which masses can be determined always improves when considering correlations. For our choice of the error bars and the correlation parameter, the precision in mass improves by up to 60\%, \ie masses can be determined almost twice as precisely. This also means that the luminosities of stars can be predicted more precisely and hence also the spectroscopic distance. Furthermore, the age can be determined significantly more precisely (up to 60\%) in those areas in the Kiel diagram where the isochrones tend to be parallel to the error ellipse. Close to the ZAMS the isochrones are almost perpendicular to the rotated error ellipses and the precision in age lowers. 
Overall, the changes in precision in mass and age are larger for our stars in the Kiel diagram than for stars in the HR diagram because of our choice of stronger correlations in effective temperature and surface gravity rather than of effective temperature and luminosity.

\subsubsection{Combined HR and Kiel diagrams}\label{sec:impact-both-diagrams}

\begin{figure*}
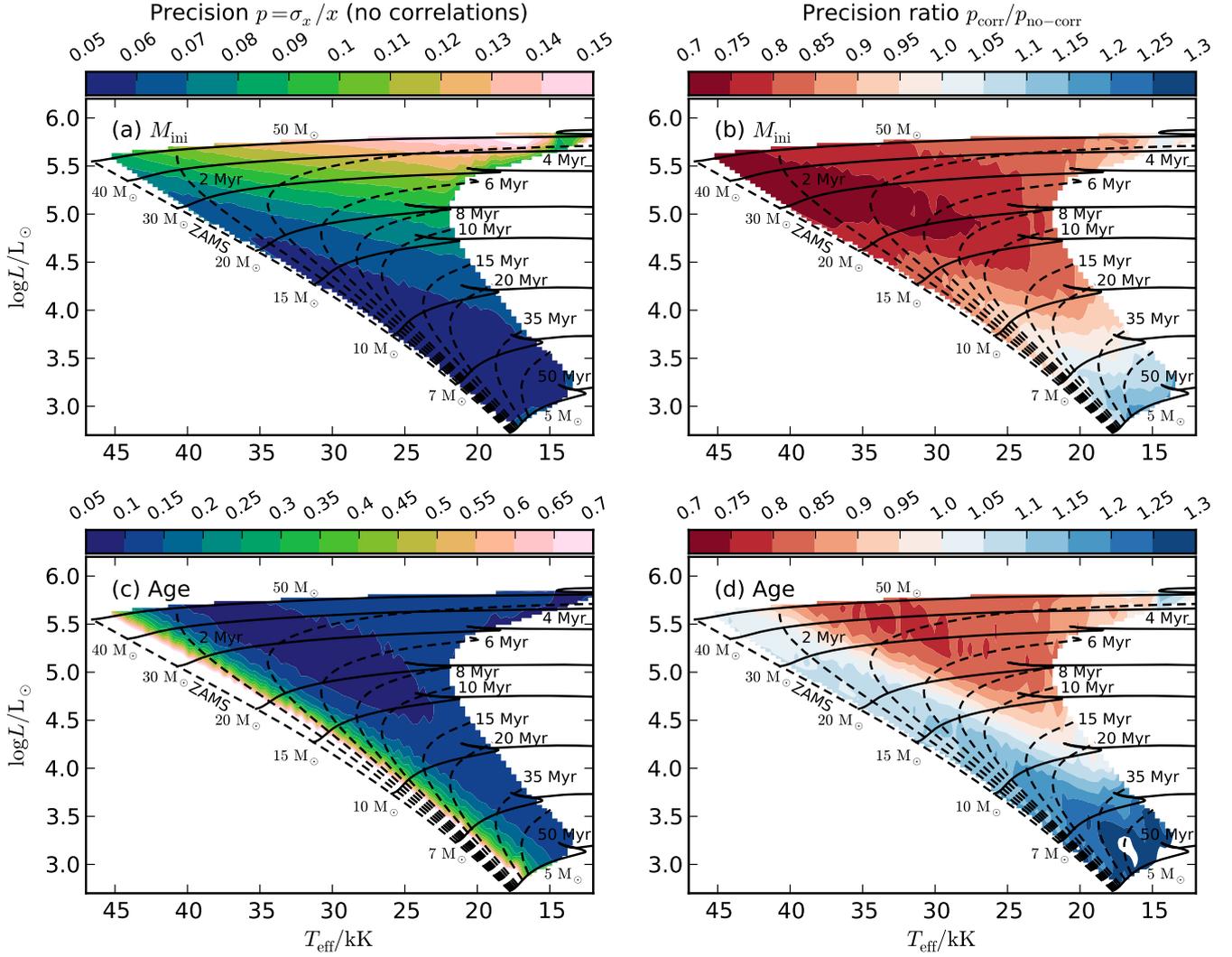

\begin{centering}
\includegraphics[width=18cm]{{{prec-ratio-map-tlg_dTeff-1000_dlogL-0.10_dlogg-0.10-prec}}}
\par\end{centering}
\caption{Same as Figs.~\ref{fig:prec-hrd} and~\ref{fig:prec-kiel} but for stars with known positions in the HR and Kiel diagrams. The $1\sigma$ uncertainties are $1000\,\mathrm{K}$ in effective temperature, $0.1\,\mathrm{dex}$ in logarithmic luminosity, and $0.1\,\mathrm{dex}$ in logarithmic surface gravity. The correlation parameters are the same as in Secs.~\ref{sec:impact-hr-diagram} and~\ref{sec:impact-kiel-diagram}.}
\label{fig:prec-hrd+kiel}
\end{figure*}

We now match effective temperatures, surface gravities, and luminosities simultaneously to stellar models while applying the same correlations as before. The precisions with which mass and age can be determined are shown in Fig.~\ref{fig:prec-hrd+kiel}. In general it is expected that the mass and age precisions improve (or at least remain the same) when matching all three observables against stellar models and not just two observables because there is more information to better constrain fundamental stellar parameters. However, the improvements may not be large and are only significant whenever the third observable adds valuable information. In Fig.~\ref{fig:implications-examples} the chosen error ellipse for Star~A and~B in the Kiel diagram span a wider range of initial masses and ages than in the HR diagram. This is reflected in the mass and age precisions from stars in the HR and Kiel diagrams where we find that both mass and age can be inferred to a higher precision from the HR diagram because of our choice of error bars (Figs.~\ref{fig:prec-hrd} and~\ref{fig:prec-kiel}). Thus, the precisions in mass improve only marginally when matching all three observables against stellar models and those in age improve significantly only when the third observable adds valuable new information, \eg if the density of isochrones is high (close to the ZAMS).

The change of the precisions when incorporating correlations is a combination of the changes discussed in Secs.~\ref{sec:impact-hr-diagram} and~\ref{sec:impact-kiel-diagram}. For example, the mass precision is lower for stars in the HR diagram (Fig.~\ref{fig:prec-hrd}b) and improves for stars in the Kiel diagram (Fig.~\ref{fig:prec-kiel}b). In stars more massive than about $7\,\msun$, the improvement due to correlations in effective temperature and gravity outweigh the lowering because of correlations in effective temperature and luminosity, while this is the opposite in stars less massive than $7\,\msun$ (Fig.~\ref{fig:prec-hrd+kiel}b). A similar situation is found for the age precisions (Fig.~\ref{fig:prec-hrd+kiel}d): correlations in effective temperatures and luminosities improve the age precision for stars close to the ZAMS, while age precisions are lower for stars that are farther away (Fig.~\ref{fig:prec-hrd}d). The opposite trend is found for correlations in effective temperatures and surface gravities (Fig.~\ref{fig:prec-kiel}d). Altogether the effects on the age precisions tend to compensate, but the correlation in effective temperature and gravity dominates and therefore outweighs the changes in the precisions.

\subsection{Most likely inferred initial masses and ages}\label{sec:most-likely-parameters}

\begin{figure*}
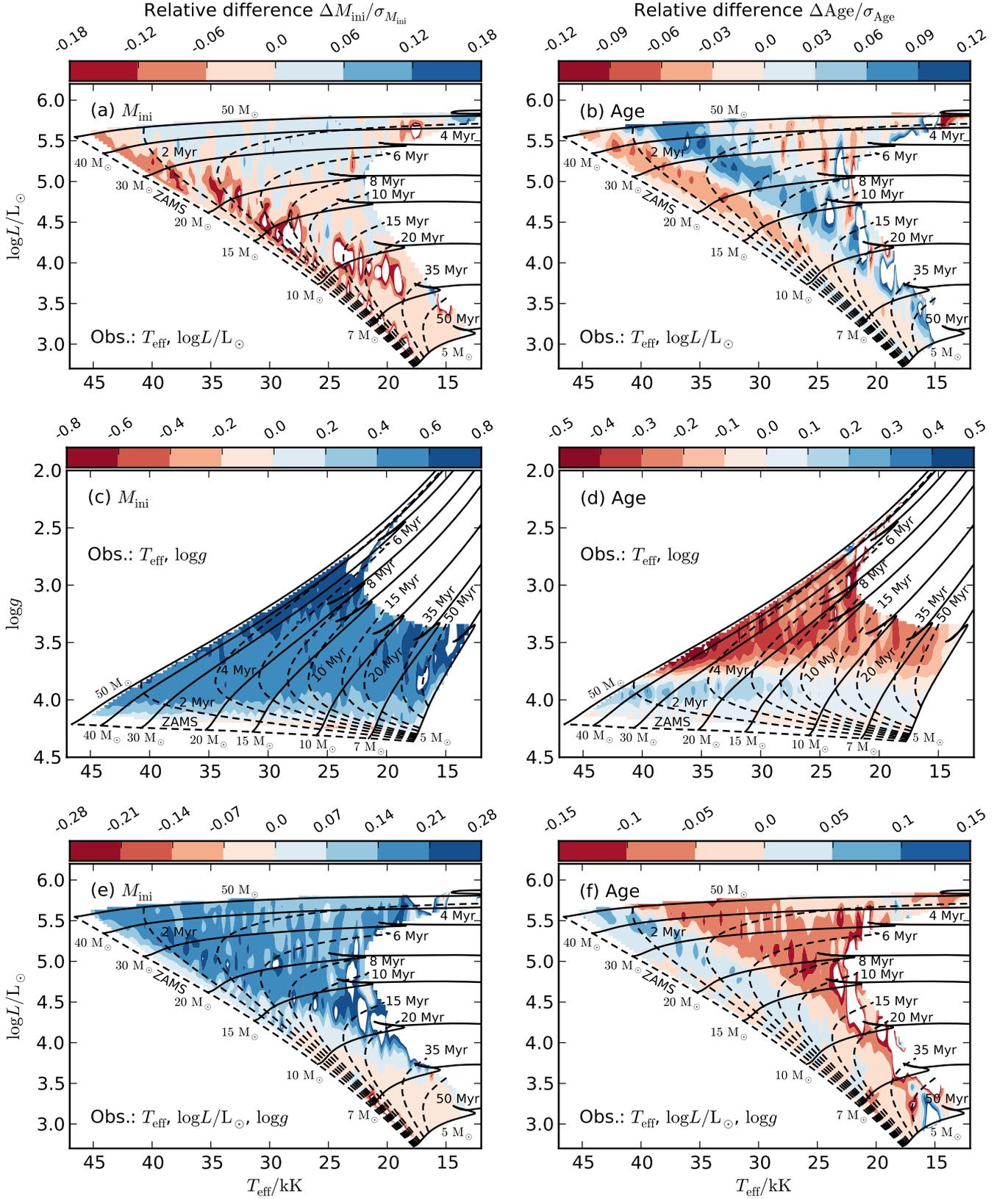

\begin{centering}
\includegraphics[width=18cm]{{{diff-map-tl_dTeff-1000_dlogL-0.10_dlogg-0.10}}}
\includegraphics[width=18cm]{{{diff-map-tg_dTeff-1000_dlogL-0.10_dlogg-0.10}}}
\includegraphics[width=18cm]{{{diff-map-tlg_dTeff-1000_dlogL-0.10_dlogg-0.10}}}
\par\end{centering}
\caption{Relative differences in the inferred most likely initial mass (left panels) and age (right panels) when taking correlations into account. In the top panels (a) and (b) we show the relative differences for stars with known effective temperatures and luminosities, in the middle panels (c) and (d) for stars with known effective temperatures and surface gravities, and in the bottom panels (e) and (f) for stars with known effective temperatures, luminosities, and surface gravities. The differences are defined with respect to the most likely parameters when neglecting correlations, \ie $(x_\mathrm{corr}-x_\mathrm{no-corr})/\sigma_\mathrm{no-corr}$ where $x$ is the parameter and $\sigma$ the $1\sigma$ uncertainty. The indices ``corr'' and ``no-corr'' show that correlations have been taken into account and neglected, respectively. The different ranges in the relative differences in each panel should be noted.}
\label{fig:most-likely-parameters}
\end{figure*}

Not only the precision, but also the most likely inferred stellar parameters are modified when accounting for correlations. To judge the importance of these changes, we have to compare them to the inferred error bars. To this end, we systematically map the relative differences $(x_\mathrm{corr}-x_\mathrm{no-corr})/\sigma_\mathrm{no-corr}$ of inferred initial masses and ages in Fig.~\ref{fig:most-likely-parameters} (the indices ``corr'' and ``no-corr'' refer to the case when taking into account and neglecting correlations, respectively, and $\sigma$ is the $1\sigma$ uncertainty of parameter $x$). As before, we consider the three cases separately when the positions of stars in the HR diagram, Kiel diagram, and both diagrams are known with $1\sigma$ uncertainties of $1000\,\mathrm{K}$, $0.1\,\mathrm{dex}$, and $0.1\,\mathrm{dex}$ in effective temperature, luminosity, and surface gravity, respectively. In all cases the differences are within the $1\sigma$ uncertainties of the stellar parameters, but still result in systematic biases when neglecting correlations and may thus be particularly important when considering samples of stars. 

As is evident from Figs.~\ref{fig:most-likely-parameters}a and~\ref{fig:most-likely-parameters}b, the differences in initial mass and age for our stars in the HR diagram are of the order of $\pm5\%$ of the $1\sigma$ error bars and at most about 20\% and 15\%, respectively. The correlations are stronger for our stars in the Kiel diagram, leading to larger relative differences in the most likely inferred masses and ages (Figs.~\ref{fig:most-likely-parameters}c and~\ref{fig:most-likely-parameters}d). Masses are systematically more massive by about 50\% of the $1\sigma$ error bars. Ages can be systematically younger or older depending on the relative orientation of the error ellipse and isochrones. On average, ages change by about $+10\%$ and $-25\%$ of $1\sigma$ uncertainties. In the worst case, masses are underestimated by $0.8\sigma$ and ages over- and underestimated by $0.5\sigma$ and $0.3\sigma$, respectively, showing the importance of correlations in this case. Matching the positions of stars in the HR and Kiel diagram simultaneously to stellar models results in slightly larger changes than what was found for stars in the HR diagram and smaller changes than for stars in the Kiel diagram. The differences are of the order of $15\%$ in mass and $8\%$ in age with respect to their corresponding $1\sigma$ uncertainties (Figs.~\ref{fig:most-likely-parameters}e and~\ref{fig:most-likely-parameters}f).

\section{Conclusion}\label{sec:conclusion}

In this paper, we include the covariance matrix in the likelihood model of the Bayesian code \bonnsai \citep{2014A&A...570A..66S} to facilitate the use of correlated observables when matching observations of stars against stellar models. Because correlations are typically not published and often not available, we derive a parametrisation of the covariance matrix that requires conventional observables including their uncertainties and additionally a correlation parameter that describes how two observables co-vary and that only depends on the method used to determine the observables. The advantage of this parametrisation is that correlations can be easily incorporated and may even be used when information on correlations are not published with the data but are known in general.

Correlations modify the likelihood function and, in our case, the likelihood function has the shape of a rotated ellipse. Because of the rotation and change in the shape of the error ellipses, the likelihood function covers different parts of the parameter space of the observables, \eg in the HR diagram, compared to the case when correlations are neglected. As a result of neglecting correlations, the most likely value of inferred stellar parameters such as mass and age can be systematically biased and the inferred error bars, \ie the precision with which stellar parameters can be determined, can be under- or overestimated. We show that the importance of correlations depends on the interplay between the strength of correlations and the conventional error bars.

In our example of OB stars with effective temperatures and surface gravities being determined from atmosphere modelling, the correlations are significant and we find that the precision with which masses can be derived improves by about a factor of 2. At the same time the precision in age decreases over a wide range of effective temperatures and surface gravities. We also find that initial masses are systematically underestimated on average by $0.5\sigma$ and ages often systematically overestimated when neglecting correlations for these stars in the Kiel diagram. In all of the cases, the reason for the differences is the change in the orientation of the likelihood model with respect to the stellar tracks and isochrones.

A likelihood model that takes correlations properly into account is a better approximation of the data. The reliability and robustness of inferred fundamental stellar parameters is therefore always enhanced, manifesting itself in systematic changes in the precision and most likely values of inferred stellar parameters. This fosters the importance of taking correlations properly into account when approaching an era of precision stellar astrophysics with current and upcoming surveys such as \gaia, and whenever robust error bars are essential.

\begin{acknowledgements}
We thank J.~Puls for helpful discussions regarding correlations from stellar atmosphere modelling, and an anonymous referee for carefully reading our manuscript and making constructive suggestions. FRNS thanks the Hintze Family Charitable Foundation and Christ Church College for his Hintze and postdoctoral research fellowships.
\end{acknowledgements}

\appendix

\section{Influence of prior distributions}\label{sec:influence-priors}

In a Bayesian analysis, not only does the likelihood model influence inferred model parameters (precisions and most likely values), but also the prior functions. In our default statistical model, we use a Salpeter IMF as initial mass prior distribution, a Gaussian as initial rotational velocity prior distribution, and a uniform age prior distribution, which are adequate descriptions of prior knowledge of massive, Milky Way stars.

\begin{figure*}
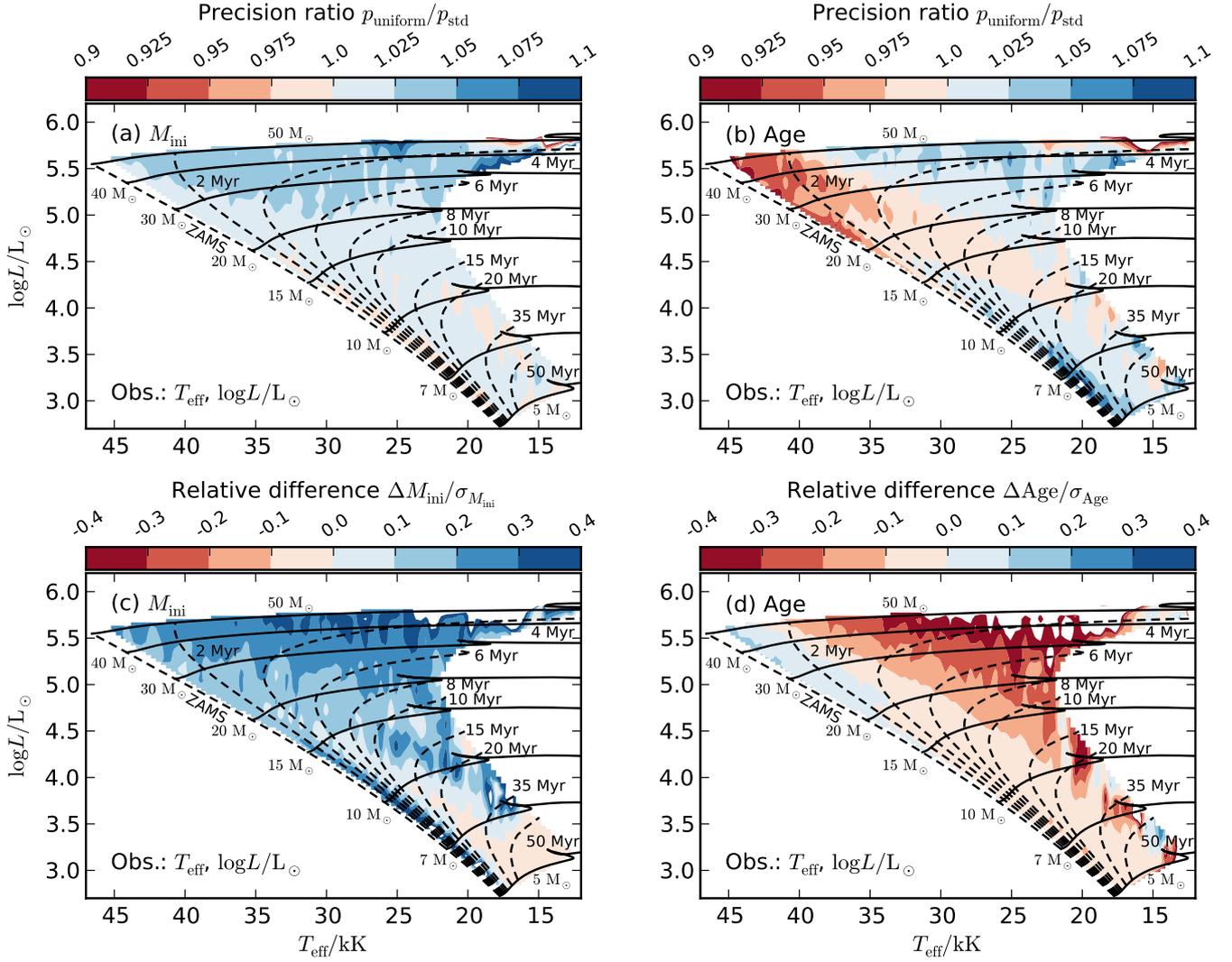

\begin{centering}
\includegraphics[width=18cm]{{{prior-comparisons-tl-mini-prior_dTeff-1000_dlogL-0.10_dlogg-0.10}}}
\par\end{centering}
\caption{Changes in the precisions $p=\sigma_x/x$ (top row) and most likely values (bottom row) of initial mass (left column) and age (right column) when assuming a uniform $M_\mathrm{ini}$-prior instead of the Salpeter IMF prior distribution applied in our default statistical model (the changes in most likely values are defined analogously to those in Fig.~\ref{fig:most-likely-parameters}, \ie $\Delta x = x_\mathrm{uniform}-x_\mathrm{std}$; $x_\mathrm{uniform}$ and $x_\mathrm{std}$, and $p_\mathrm{uniform}$ and $p_\mathrm{std}$ denote the most likely values and precisions for a uniform $M_\mathrm{ini}$-prior distribution and our standard prior choice, respectively). The observables are effective temperature and luminosity, and the precision ratios and relative differences are with respect to the precisions and most likely values when neglecting correlations (cf. Figs.~\ref{fig:prec-hrd}, \ref{fig:most-likely-parameters}a, and~\ref{fig:most-likely-parameters}b). For more details, see Sec.~\ref{sec:influence-priors}.}
\label{fig:prior-comparisons-mini}
\end{figure*}

A Salpeter IMF prior distribution puts more weight on lower mass stars. Any $M_\mathrm{ini}$-prior distribution influences the inference of initial masses most strongly in a region of the observational parameter space where tracks of different initial mass are closest to each other. For example, in the HR diagram stellar tracks are closer to each other when the initial mass is higher. Changing the $M_\mathrm{ini}$-prior distribution will therefore have a greater influence on higher mass stars than on lower mass stars (assuming constant error bars). The main influence of the $M_\mathrm{ini}$-prior distribution is to change the most likely values of inferred initial masses and ages; for example, replacing the Salpeter IMF prior by a uniform $M_\mathrm{ini}$-prior distribution for stars in the HR diagram changes the precision with which mass and age can be determined by up to $\pm5\%$, and shifts all masses to higher values and ages to lower values because of the anti-correlation of mass and age (more massive stars have shorter lifetimes). The shift is strongest for the most massive stars and can be up to $+0.4\sigma$ in mass and $-0.4\sigma$ in age (Fig.~\ref{fig:prior-comparisons-mini}).

\begin{figure*}
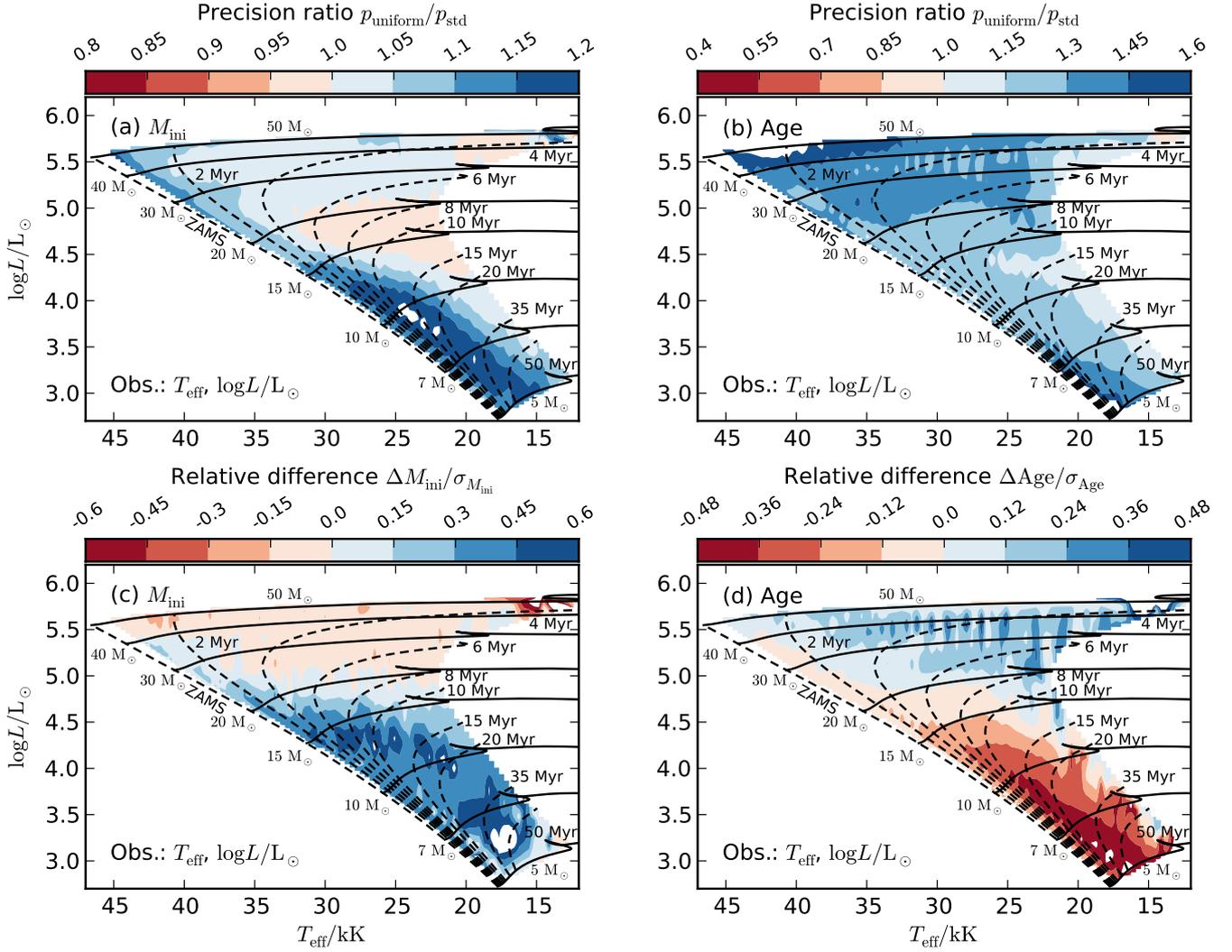

\begin{centering}
\includegraphics[width=18cm]{{{prior-comparisons-tl-vini-prior_dTeff-1000_dlogL-0.10_dlogg-0.10}}}
\par\end{centering}
\caption{Same as Fig.~\ref{fig:prior-comparisons-mini} but applying a uniform $v_\mathrm{ini}$-prior instead of the Gaussian $v_\mathrm{ini}$-prior distribution used in our default statistical model. As $M_\mathrm{ini}$-prior distribution, we use a Salpeter IMF here. For more details, see Sec.~\ref{sec:influence-priors}.}
\label{fig:prior-comparisons-vini}
\end{figure*}

Rotation can significantly change stellar models. In general, rotation ``blurs'' tracks of the same initial mass in the HR diagram such that a wider variety of initial masses can reach the same effective temperatures and luminosities. Giving more weight to a broader range of initial rotational velocities, for example by choosing a greater width of the Gaussian $v_\mathrm{ini}$-prior distribution or by even using a uniform prior distribution, a wider range of models becomes likely to explain the same position of stars in the HR diagram, hence lowering the precisions with which mass and age can be inferred. Furthermore, rotating stars can be either more or less luminous than non-rotating stars depending on their mass and evolutionary stage. This is due to a balance of rotationally induced chemical mixing making stars more luminous and centrifugal forces making them less luminous (on the ZAMS, rotating stellar models are therefore always less luminous because mixing has had no effect yet). In the Milky Way models of \citet{2011A&A...530A.115B}, (rapidly) rotating stars more massive than about $20\,\msun$ can become considerably more luminous than their non-rotating counterparts during the MS evolution. Giving more weight to rapid rotators by changing (broadening) the $v_\mathrm{ini}$-prior distribution will therefore shift the most likely mass to smaller values for $M_\mathrm{ini}\gtrsim 20\,\msun$ because less massive, rapid rotators reaching the same effective temperatures and luminosities as slowly rotating, more massive stars are no longer disfavoured as much as before by the prior distribution (or are not disfavoured at all for a uniform prior distribution). Analogously, the most likely masses shift to larger values for $M_\mathrm{ini}\lesssim 20\,\msun$, and the most likely ages will be older for $M_\mathrm{ini}\gtrsim 20\,\msun$ and younger for $M_\mathrm{ini}\lesssim 20\,\msun$. Close to the ZAMS, inferred masses always shift to larger values and ages to lower values. In the most extreme (and unrealistic) case of using a uniform $v_\mathrm{ini}$-prior distribution, the precisions of inferred masses and ages is reduced by up to $20\%$ and $60\%$, respectively. At the same time, the most likely masses and ages can change by up to $\pm 0.5\text{--}0.6\sigma$ (Fig.~\ref{fig:prior-comparisons-vini}).

The changes in precision and most likely mass and age quoted above can be regarded as upper limits because they originate from changing either the $M_\mathrm{ini}$- or the $v_\mathrm{ini}$-prior distribution by a maximum extent such that all initial masses or initial rotational velocities are a priori equally probable. This situation corresponds to going from a simple maximum-likelihood approach (uniform prior distributions) to a full Bayesian approach. The influence of changing prior distributions on inferred model parameters are qualitatively the same for stars in the Kiel diagram, but differ quantitatively.

\bibliographystyle{aa}

\end{document}